\documentclass[sigconf]{acmart}

\AtBeginDocument{%
  \providecommand\BibTeX{{%
    \normalfont B\kern-0.5em{\scshape i\kern-0.25em b}\kern-0.8em\TeX}}}

\usepackage{booktabs}
\usepackage{array}
\usepackage{balance} 
\usepackage{lipsum}
\usepackage{multirow}
\usepackage{booktabs}
\usepackage{amsmath}
\usepackage{subfigure}
\usepackage{algorithm}  
\usepackage{algorithmicx}  
\usepackage{algpseudocode}  
\usepackage{amsmath}  
\usepackage{enumitem}
\usepackage{tabularx}
\usepackage[utf8]{inputenc}
\usepackage[english]{babel}
\usepackage{amsthm}
\usepackage{bm}
\newcommand{\ie}{\emph{i.e., }}
\newcommand{\eg}{\emph{e.g., }}
\newcommand{\etal}{\emph{et al. }}

\newcommand{\wrt}{\emph{w.r.t. }}

\newlength\myindent
\usepackage[normalem]{ulem}
\useunder{\uline}{\ul}{}

\floatname{algorithm}{Algorithm}

\copyrightyear{2022}
\acmYear{2022}
\setcopyright{acmcopyright}\acmConference[SIGIR '22]{Proceedings of the 45th International ACM SIGIR Conference on Research and Development in Information Retrieval}{July 11--15, 2022}{Madrid, Spain}
\acmBooktitle{Proceedings of the 45th International ACM SIGIR Conference on Research and Development in Information Retrieval (SIGIR '22), July 11--15, 2022, Madrid, Spain}
\acmPrice{15.00}
\acmDOI{10.1145/3477495.3532075}
\acmISBN{978-1-4503-8732-3/22/07}

\begin{document}

\title{User-controllable Recommendation Against Filter Bubbles}

\author{Wenjie Wang$^{1}$, Fuli Feng$^{2*}$, Liqiang Nie$^{3}$, and Tat-Seng Chua$^1$}
\def\authors{Wenjie Wang, Fuli Feng, Liqiang Nie, and Tat-Seng Chua}
\affiliation{
\institution{$^1$Sea-NExT Joint Lab, National University of Singapore,\\ $^2$University of Science and Technology of China, $^3$Shandong University}
\country{}
}
\email{{wenjiewang96, fulifeng93, nieliqiang}@gmail.com, dcscts@nus.edu.sg}
\thanks{$*$ Corresponding author: Fuli Feng. This research is supported by the Sea-NExT Joint Lab, the National Natural Science Foundation of China (No. U1936203), and the Major Basic Research Project of Natural Science Foundation of Shandong Province (No. ZR2021ZD15).}

\renewcommand{\shortauthors}{Wenjie Wang, Fuli Feng, Liqiang Nie, and Tat-Seng Chua}

\begin{abstract}

Recommender systems usually face the issue of filter bubbles: over-recommending homogeneous items based on user features and historical interactions. Filter bubbles will grow along the feedback loop and inadvertently narrow user interests. Existing work usually mitigates filter bubbles by incorporating objectives apart from accuracy such as diversity and fairness. However, they typically sacrifice accuracy, hurting model fidelity and user experience. Worse still, users have to passively accept the recommendation strategy and influence the system in an inefficient manner with high latency, \eg keeping providing feedback (\eg like and dislike) until the system recognizes the user intention.

This work proposes a new recommender prototype called \textit{User-Controllable Recommender System} (UCRS), which enables users to actively control the mitigation of filter bubbles. Functionally, 1) UCRS can alert users if they are deeply stuck in filter bubbles. 2) UCRS supports four kinds of control commands for users to mitigate the bubbles at different granularities. 3) UCRS can respond to the controls and adjust the recommendations on the fly. The key to adjusting lies in blocking the effect of out-of-date user representations on recommendations, which contains historical information inconsistent with the control commands. As such, we develop a causality-enhanced \textit{User-Controllable Inference} (UCI) framework, which can quickly revise the recommendations based on user controls in the inference stage and utilize counterfactual inference to mitigate the effect of out-of-date user representations. Experiments on three datasets validate that the UCI framework can effectively recommend more desired items based on user controls, showing promising performance \wrt both accuracy and diversity. 

\end{abstract}



\maketitle

\begin{figure}[t]
\setlength{\abovecaptionskip}{0cm}
\setlength{\belowcaptionskip}{-0.2cm}
\centering
\includegraphics[scale=0.42]{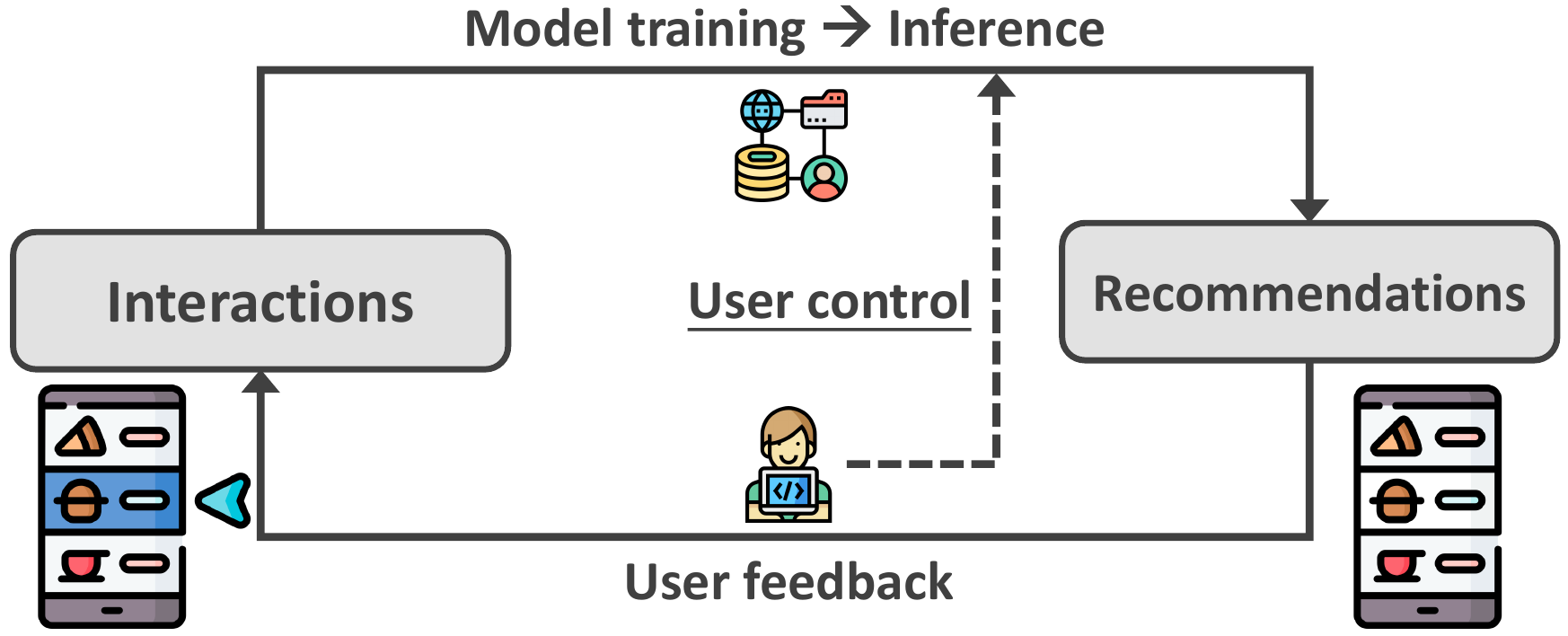}
\caption{Illustration of feedback loop and user controls. User feedback passively affects the recommendation strategy while user controls can directly adjust the strategy.}
\label{fig:feedback_loop}
\vspace{-0.08cm}
\end{figure}

\section{Introduction}
\label{sec:introduction}

Recommender systems become increasingly important to provide personalized information filtering services in this information explosion era~\cite{rendle2010factorization}. A \textit{de facto} standard for building recommender systems is mining user interests from user features (\eg gender and age) and historical interactions (\eg click). Due to merely fitting the data, recommender systems typically face filter bubble issues: continually recommending many homogeneous items, isolating users from diverse contents~\cite{wang2021deconfounded, pariser2011filter, Ge2020Understanding}.
For example, if a user has clicked many micro-videos to learn making coffee, the system may continuously recommend similar micro-videos from different uploaders, occupying the opportunities of other informative videos such as hot news. Worse still, due to the feedback loop as shown in Figure \ref{fig:feedback_loop}, filter bubbles might gradually become severer, narrow users' interests, and even intensify the segregation between users~\cite{pariser2011filter}. In the long-term, filter bubbles will decrease user activeness and item originality, hurting the ecosystem. 
Therefore, it is essential to mitigate filter bubbles.

Towards the goal, existing studies prevent the recommendations from merely fitting historical interactions by incorporating additional objectives. For instance, 
1) diversity~\cite{zheng2021dgcn, chandar2013preference}, which pushes the recommendation list to cover more item categories; 
2) fairness~\cite{Morik2020Controlling, biega2018equity}, which pursues fair exposure opportunities over item categories;
and 3) calibration~\cite{steck2018calibrated, wang2021deconfounded}, which ensures that the recommendation list exhibits the same distribution over item categories as the user's history. 
However, these methods typically make the trade-off across multiple objectives, sacrificing accuracy and even degrading user experience~\cite{steck2018calibrated, zheng2021dgcn}. Moreover, in the feedback loop, users passively adjust the recommendations by user feedback (\eg click, like, and dislike), which is inefficient and inadequate because users need to constantly provide user feedback until the system recognizes users' intention. 

We argue that users have the right to decide whether to mitigate filter bubbles and choose which bubble to mitigate. To this end, 
we conceptually propose a new prototype called \textit{User-Controllable Recommender System} (UCRS) with three main considerations: 1) the system has the responsibility to remind users if they are stuck in filter bubbles; 2) the system should provide various commands to fully support users' control intentions; and 3) the system should respond to the controls on the fly. UCRS achieves the three objectives with three additional functions beyond conventional recommender systems. 
\begin{itemize}[leftmargin=*]
    \item \textbf{Filter bubble alert.} We define several metrics to measure the strength of filter bubbles. 
    With these metrics, \eg presented as a system notification, we aim to let users understand the status of filter bubbles and decide whether to mitigate the bubbles.
    
    \item \textbf{Control commands.} We suggest user controls at two levels regarding either a user or item feature. At the fine-grained level, UCRS supports the commands to increase the items \wrt a specified user or item feature, such as ``more items liked by \textit{young} users'' and ``more items in a target \textit{category} (\eg action movies)''. Noticing that users may not intend to specify the target group, UCRS also supports commands at the coarse-grained level, \eg ``no bubble \wrt \textit{my age}'' and ``no bubble \wrt \textit{item category}''.

    \item \textbf{Response to user controls.}
    Once receiving control commands, UCRS adjusts the recommendations by incorporating the commands into recommender inference\footnote{UCRS cannot respond to user controls timely by model retraining or fine-tuning. The computation cost is also unaffordable.}. This is however non-trivial because some out-of-date user representations learned from historical interactions have encoded the preference information leading to filter bubbles. Thus such user representations can still cause homogeneous recommendations. 

\end{itemize}

To tackle the challenges, we propose a causality-enhanced \textit{User-Controllable Inference} (UCI) framework, which inspects the generation procedure of recommendations from a causal view and leverages counterfactual inference to mitigate the effect of out-of-date user representations. Specifically, UCI imagines a counterfactual world where out-of-date user representations are discarded, and estimates their effects as the difference between factual and counterfactual worlds. After deducting such effects, UCI incorporates the control command into recommender inference. As to user-feature controls, UCI revises the user feature specified by the control command (\eg changing age from middle age to teenager) to conduct the final inference at the two levels. As to item-feature controls, UCI adopts a user-controllable ranking policy to control the recommendations \wrt item category. Extensive experiments on three datasets validate the superiority of UCI on mitigating filter bubbles without sacrificing recommendation accuracy. 
We release the code and data at: \url{https://github.com/WenjieWWJ/UCRS}.

To sum up, the contributions of this work are threefold:
\begin{itemize}[leftmargin=*]
    \item We study a new problem of using user controls to adjust filter bubbles, and propose a user-controllable recommender prototype, emphasizing the user rights of controlling recommender systems.
    \item We propose the UCI framework, which can mitigate the effect of out-of-date user representations via counterfactual inference and perform real-time adaptation to four kinds of user controls. 
    \item We define several metrics to measure filter bubbles and conduct extensive experiments on three datasets, validating the effectiveness of UCI in maintaining the accuracy and mitigating filter bubbles by following user controls. 
\end{itemize}

\section{Related Work}
\label{sec:related_work}


\vspace{3pt}
\noindent$\bullet$ \textbf{Filter bubbles in recommendation.}
Although recommender systems have achieved great success in the past years~\cite{he2017nfm, cheng2018aspect, cheng2022feature}, the debate on filter bubbles has always attracted extensive attention~\cite{nguyen2014exploring, bakshy2015exposure}. On one side, researchers claim that recommender systems provide users with satisfying items, and might expose some items that users would never see without recommendations~\cite{nguyen2014exploring}. On the other side, many studies~\cite{pariser2011filter, sunstein2009going, dandekar2013biased} have stated that personalized recommendation would cause group polarization, where users are fragmented and the users with similar interests are grouped. Later, some work introduces filter bubbles in recommendation: users always receive similar content and gradually become isolated from diverse items~\cite{badami2018prcp, flaxman2016filter}. Moreover, due to the feedback loop~\cite{mansoury2020feedback, jiang2019degenerate}, the continual exposure of similar content will further intensify user interests over such items~\cite{chaney2018algorithmic}, leading to the issues of echo chamber~\cite{Ge2020Understanding, donkers2021dual, tommasel2021want} and ideological segregation~\cite{pariser2011filter}. In this work, we have found that filter bubbles do exist in the scenarios of content recommendation, such as movies and books. Different from previous work, our consideration is to let users decide whether to mitigate filter bubbles and directly control the recommendations. This is a big step for user engagement because it transfers the decision right from recommender platforms to users. 

\vspace{3pt}
\noindent$\bullet$ \textbf{Diversity in recommendation.}
Diversity has been widely used as one additional objective to alleviate filter bubbles~\cite{cheng2017learning}, where recommender models are encouraged to generate dissimilar items in a recommendation list~\cite{clarke2008novelty, chandar2013preference}. Generally, item similarity can be compared by various distance functions (\eg cosine similarity) and item features (\eg item category and well-trained embeddings)~\cite{ziegler2005improving}. Technically, the diversity-oriented recommendation can be divided into post-processing~\cite{carbonell1998use} and end-to-end methods~\cite{zheng2021dgcn}. The former diversifies the recommendation lists generated by some models via re-ranking~\cite{carbonell1998use, ziegler2005improving}. In contrast, the latter directly balances the objectives of accuracy and diversity during training and inference~\cite{chen2018fast}. However, existing methods simply recommend diverse items to users, and then find new item categories liked by the users. This process does not only take lots of time and user feedback, but also brings many irrelevant items~\cite{ziegler2005improving}. To solve these problems, our proposed UCRS utilizes user controls to indicate user intention and provide efficient diversification.

\vspace{3pt}
\noindent$\bullet$ \textbf{Fairness in recommendation.}
Extensive fairness-oriented work has considered encouraging equal exposure across item groups, where groups can be partitioned by item features, such as producer and category. Previous studies~\cite{li2021towards, Morik2020Controlling} usually focus on the definitions of fairness, spanning from amortized equity of attention~\cite{biega2018equity}, discounted cumulative fairness~\cite{yang2017measuring}, to multi-sided fairness~\cite{wu2021tfrom}. Besides, Steck \etal~\cite{steck2018calibrated} proposed the objective of calibration, which forces the proportion of item categories in a recommendation list to follow that of user's historical interactions~\cite{wang2021deconfounded}. Although fairness-related work is able to partly alleviate filter bubbles, they inevitably lie in the trade-off between accuracy and fairness, thus degrading the user experience. 

\vspace{3pt}
\noindent$\bullet$ \textbf{User-controllable recommendation.}
User controls can help users explicitly specify their interests and efficiently achieve recommendation adjustments~\cite{jannach2016user, tsai2018beyond, taijala2018movieexplorer}. Prior literature usually falls into two categories: user controls during the preference estimation~\cite{hijikata2012relation, knijnenburg2011each} and the controls over recommendation lists~\cite{swearingen2001beyond, schafer2002meta}. In the stage of preference estimation, recommender systems can acquire user controls by multiple ways, such as preference forms~\cite{hijikata2012relation} and interactive conversations~\cite{sun2018conversational, nie2019magic}. Once recommendations are presented, users can control by critiquing~\cite{chen2012critiquing} and interactive explanations~\cite{tan2021counterfactual, zhang2020explainable}. Although these methods can perform user controls in different stages, they ignore the advantages of user controls on alleviating filter bubbles. Besides, previous studies never consider the possible inconsistency between out-of-date user representations and user controls. As such, existing methods can serve as the interface to acquire user controls, which are then used in UCRS to mitigate filter bubbles via the causal UCI framework.

\vspace{3pt}
\noindent$\bullet$ \textbf{Causal recommendation.}
We use counterfactual inference to mitigate the effect of out-of-date user representations, which is related to causal recommender models~\cite{bonner2018causal, Thorsten2017unbiased, zou2020counterfactual}. Generally, two causal frameworks have been applied to recommendation: \textit{potential-outcome framework}~\cite{rubin2005causal} and \textit{Structural Causal Models} (SCMs)~\cite{pearl2009causality}. The former mainly leverages inverse propensity scoring~\cite{wang2018deconfounded, saito2020unbiased} and doubly robust~\cite{guo2021enhanced} to debias user feedback. SCMs typically abstract causal relationships into causal graph and estimate causal effects via intervention~\cite{wang2021deconfounded, zhang2021causal} or counterfactual inference~\cite{wang2021counterfactual, yang2021top}, which are widely used for debiasing~\cite{wang2021deconfounded}, explainable~\cite{Khanh2021Counterfactual, tan2021counterfactual}, and out-of-distribution recommendations~\cite{wang2022causal}. Nevertheless, using causality for diversity or alleviating filter bubbles receives little scrutiny.

\section{Preliminary on Filter Bubbles}
\label{sec:preliminary}

To intuitively understand filter bubbles, we conduct preliminary experiments to analyze their effects \wrt different user groups.  

\vspace{3pt}
\noindent$\bullet$ \textbf{Experimental settings.} 
We train a representative recommender model, Factorization Machine (FM)~\cite{rendle2010factorization}, on three public datasets (\ie DIGIX-Video, Amazon-Book, and ML-1M), and then collect the top-10 recommended items for each user. 
Next, to study the phenomenon of filter bubbles, we split users into groups according to two factors: user features and user interactions. Specifically, we are able to divide user groups by available user features, such as gender and age. 
Besides, different users usually have interests in different item categories (\eg romance movies), and thus we can also distinguish user groups by user interactions over item categories. For each item category, we select the users whose interaction proportion over this category is larger than a threshold (\eg $0.5$). Thereafter, we compare users' historical interactions and the recommendations generated by FM \wrt user groups. 

\begin{figure}[t]
\setlength{\abovecaptionskip}{0cm}
\setlength{\belowcaptionskip}{-0.1cm}
  \centering 
  \subfigure[Historical distributions of male and female users.]{
    \includegraphics[width=1.05in]{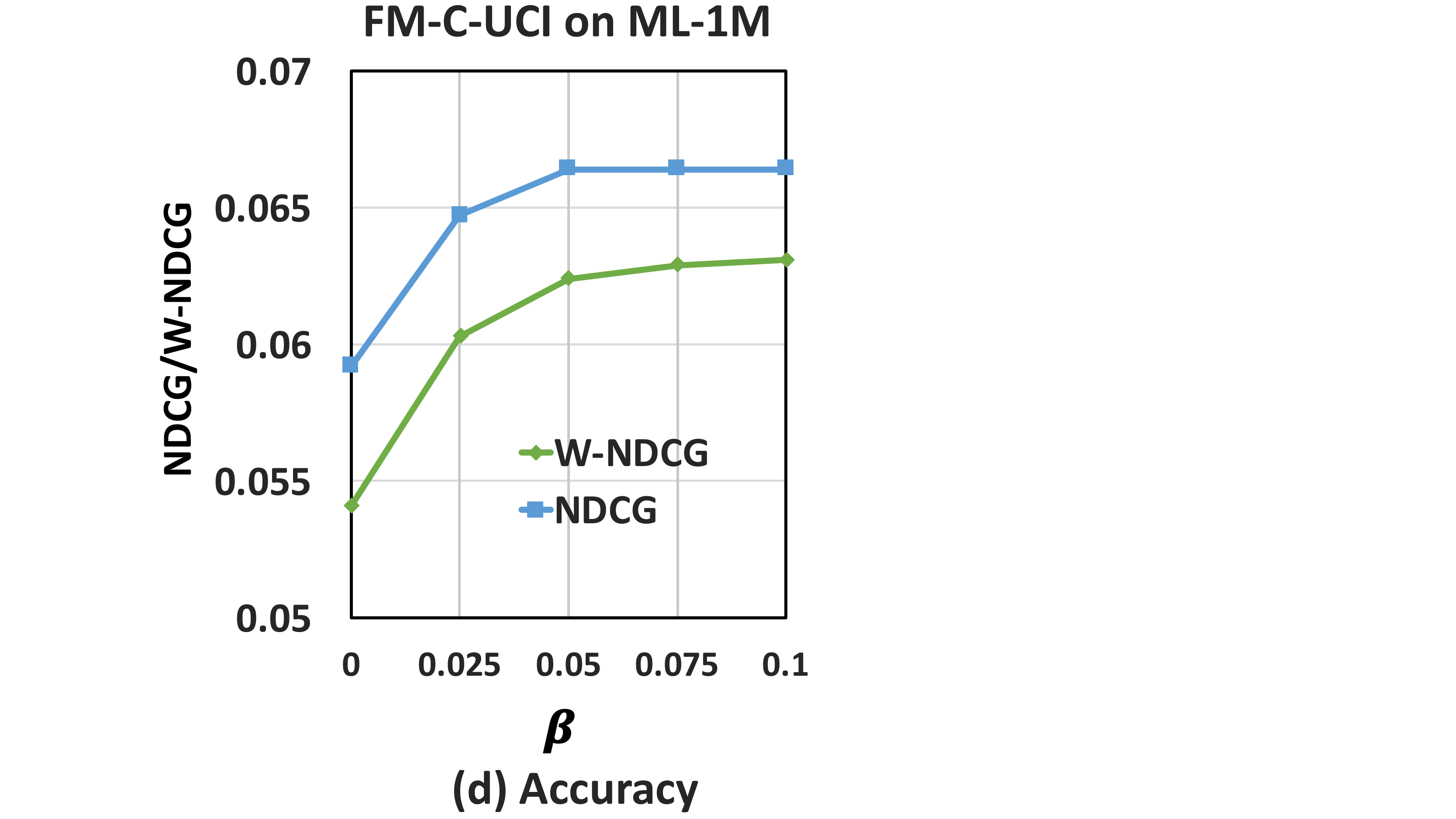}} 
  \subfigure[Category distributions of female users.]{
    \includegraphics[width=1.05in]{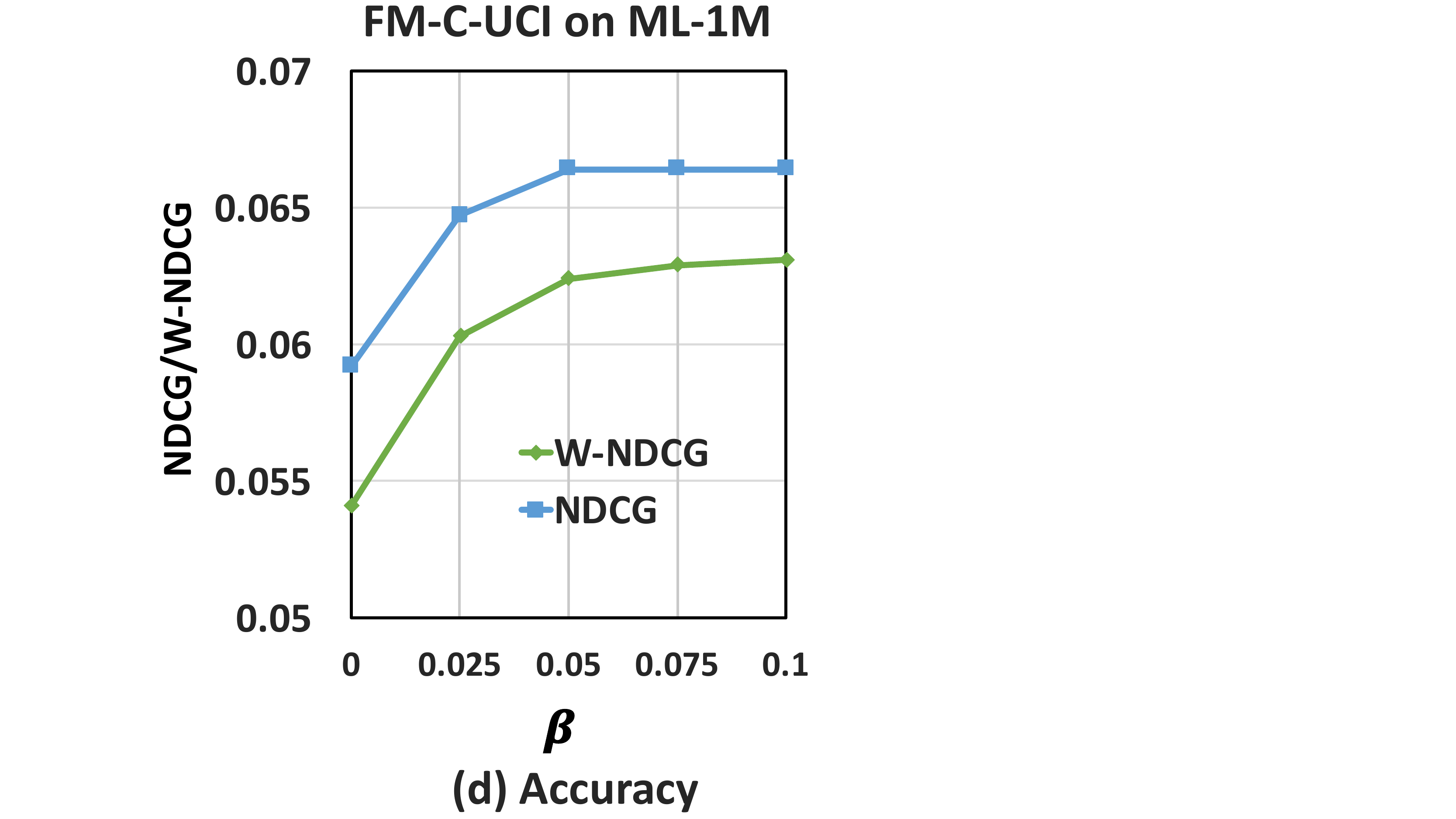}} 
  \subfigure[Category distributions of male users.]{
    \includegraphics[width=1.05in]{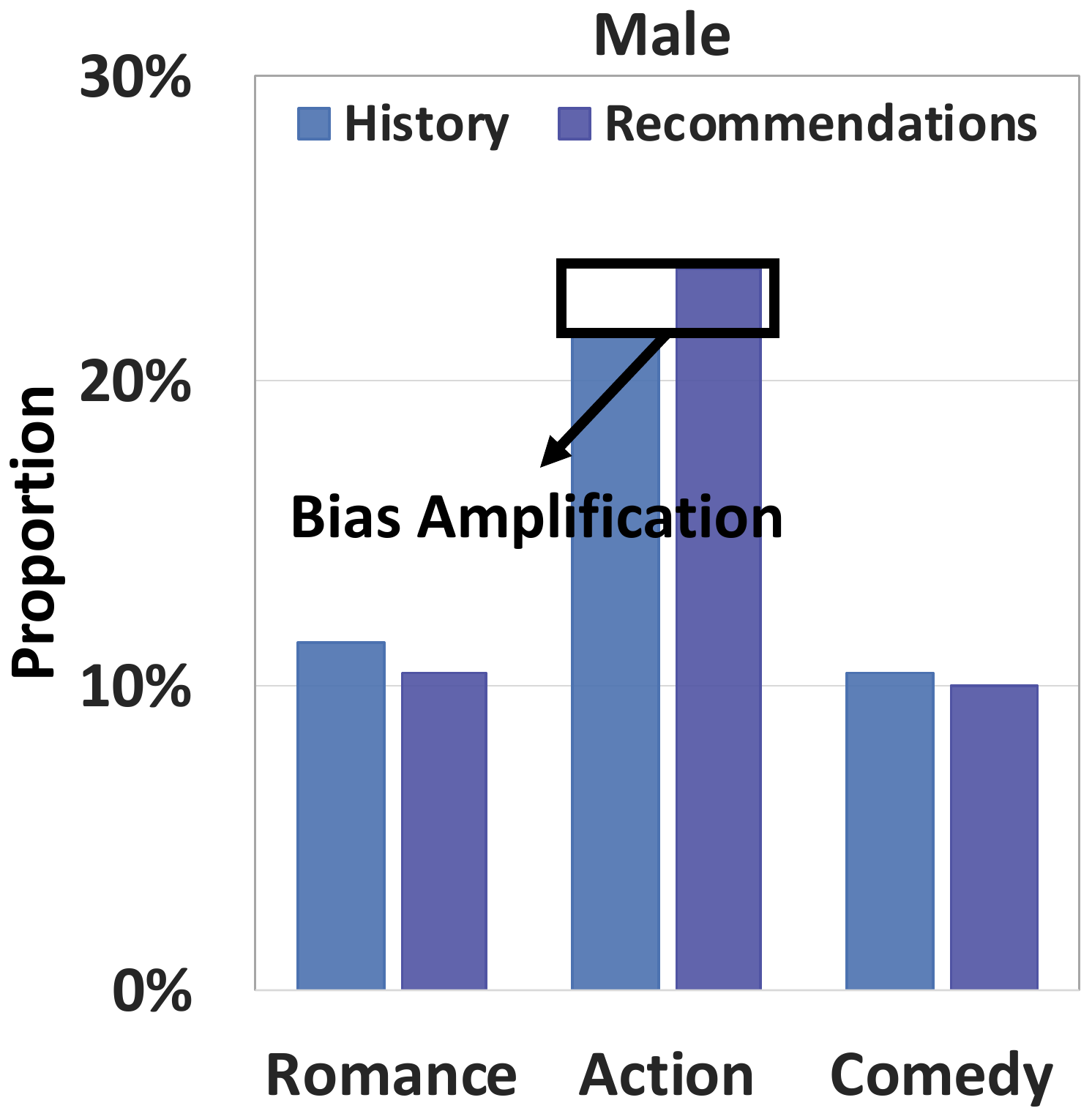}} 
  \subfigure[Category proportion in history and recommendations on Amazon-Book.]{
    \includegraphics[width=1.6in]{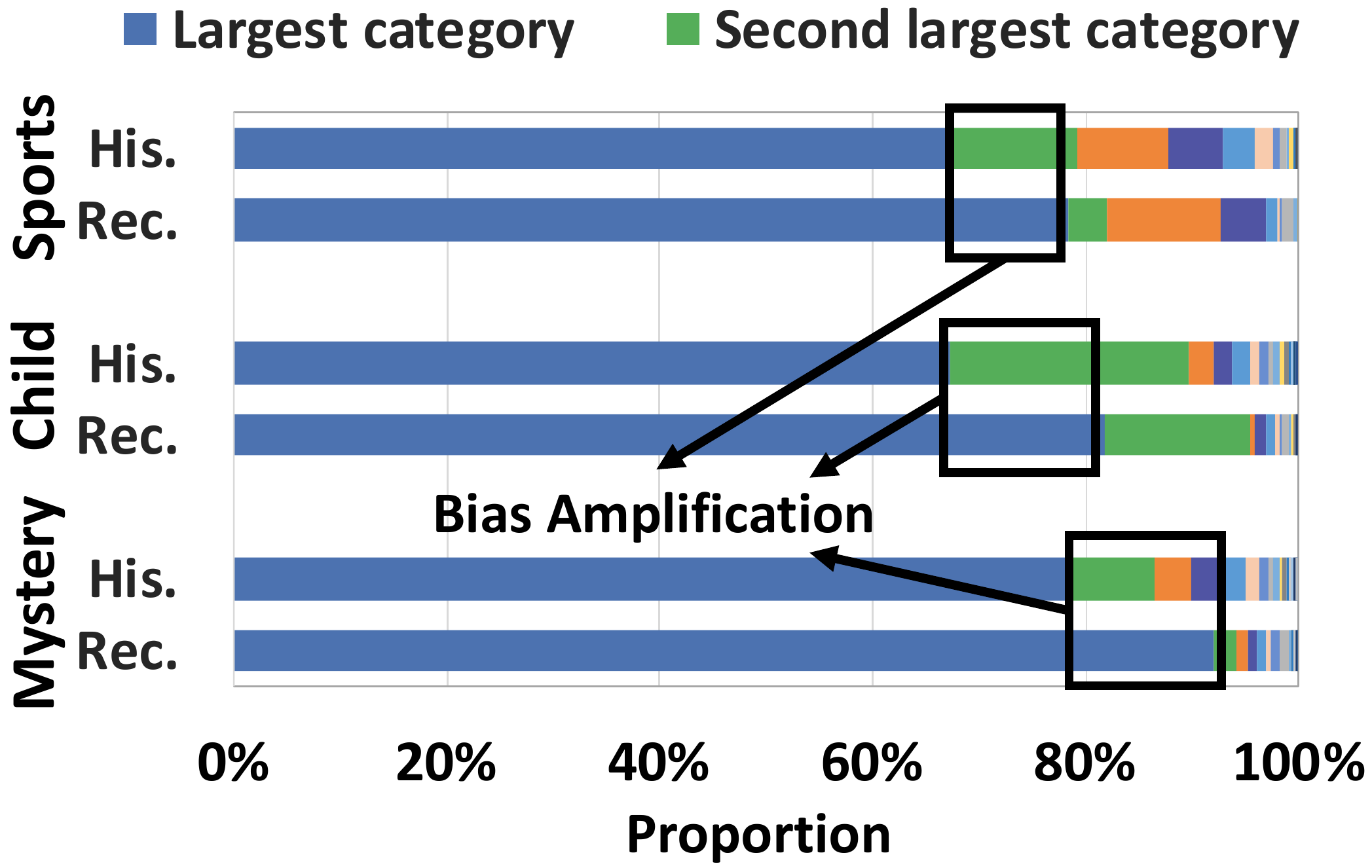}} 
  \hspace{0.01in}
  \subfigure[Category proportion in history and recommendations on ML-1M]{
    \includegraphics[width=1.6in]{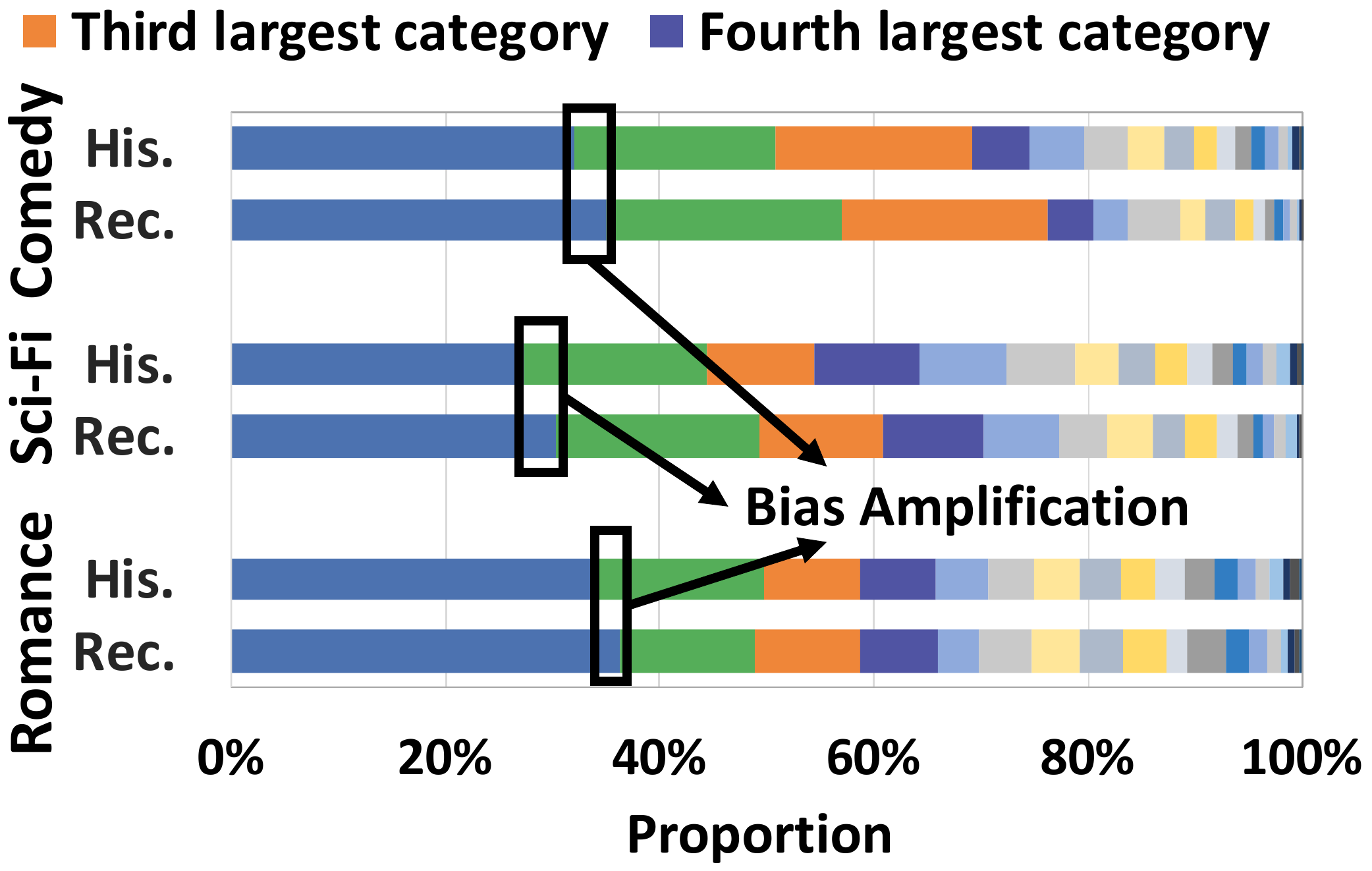}} 
  \caption{Analysis of filter bubbles and bias amplification. In Figure (d) and (e), ``His.'' and ``Rec.'' denote the historical interactions and recommendations, respectively.}
  \label{fig:preliminary}
\end{figure}

\vspace{3pt}
\noindent$\bullet$ \textbf{Analysis.}
For male and female users in DIGIX-Video, we visualize their historical distributions over top-3 item categories in Figure \ref{fig:preliminary}(a). From the figure, we can observe that male and female users express different interests in item categories. For example, as compared to females, male users prefer more action movies than romance movies. Consequently, the recommender models will inherit the biased distributions~\cite{steck2018calibrated, zhang2021causal}. As shown in Figure \ref{fig:preliminary}(b) and (c), the distribution of recommendations for male and female users is quite similar to that in the history, showing that the users will continually receive homogeneous items. Worse still, the models tend to amplify the bias and expose more historical majority categories~\cite{wang2021deconfounded} as shown in Figure \ref{fig:preliminary}(b) and (c), causing severer segregation between male and female users.

As to the user groups divided by user interactions, we present the results on Amazon-Book and ML-1M in Figure \ref{fig:preliminary}(d) and (e), respectively. The results on DIGIX-Video with similar observations are omitted to save space. From the figures, we have the following findings. 1) The largest categories in users' history are dominating the recommendation lists. Besides, as compared to Amazon-Book, the distributions on ML-1M are more diverse and the domination of majority categories is less severe. This is because most items in ML-1M have multiple categories. 2) The models usually have the bias amplification issue~\cite{wang2021deconfounded} and increase the proportions of recommended majority categories. 
Due to the bias amplification, filter bubbles will be gradually intensified, which inevitably narrow users' interests, fragment users, and lead to group segregation.

\vspace{-6pt}
\begin{center}
\fcolorbox{black}{gray!6}{\parbox{0.98\linewidth}{
\noindent$\bullet$ \textbf{Summary.}
We find that filter bubbles exist on the sides of user and item features. The bubbles \wrt item features are caused by the biased interactions over item categories. In this light, we propose the user-feature and item-feature controls correspondingly.}}
\end{center}

\section{User-controllable Recommendation}
\label{sec:method}
In this section, we first formulate the paradigm of user-controllable recommender systems, and then introduce the proposed causal UCI framework for the response to real-time user controls.

\begin{figure}[t]
\setlength{\abovecaptionskip}{0.1cm}
\setlength{\belowcaptionskip}{-0.2cm}
\centering
\includegraphics[scale=0.46]{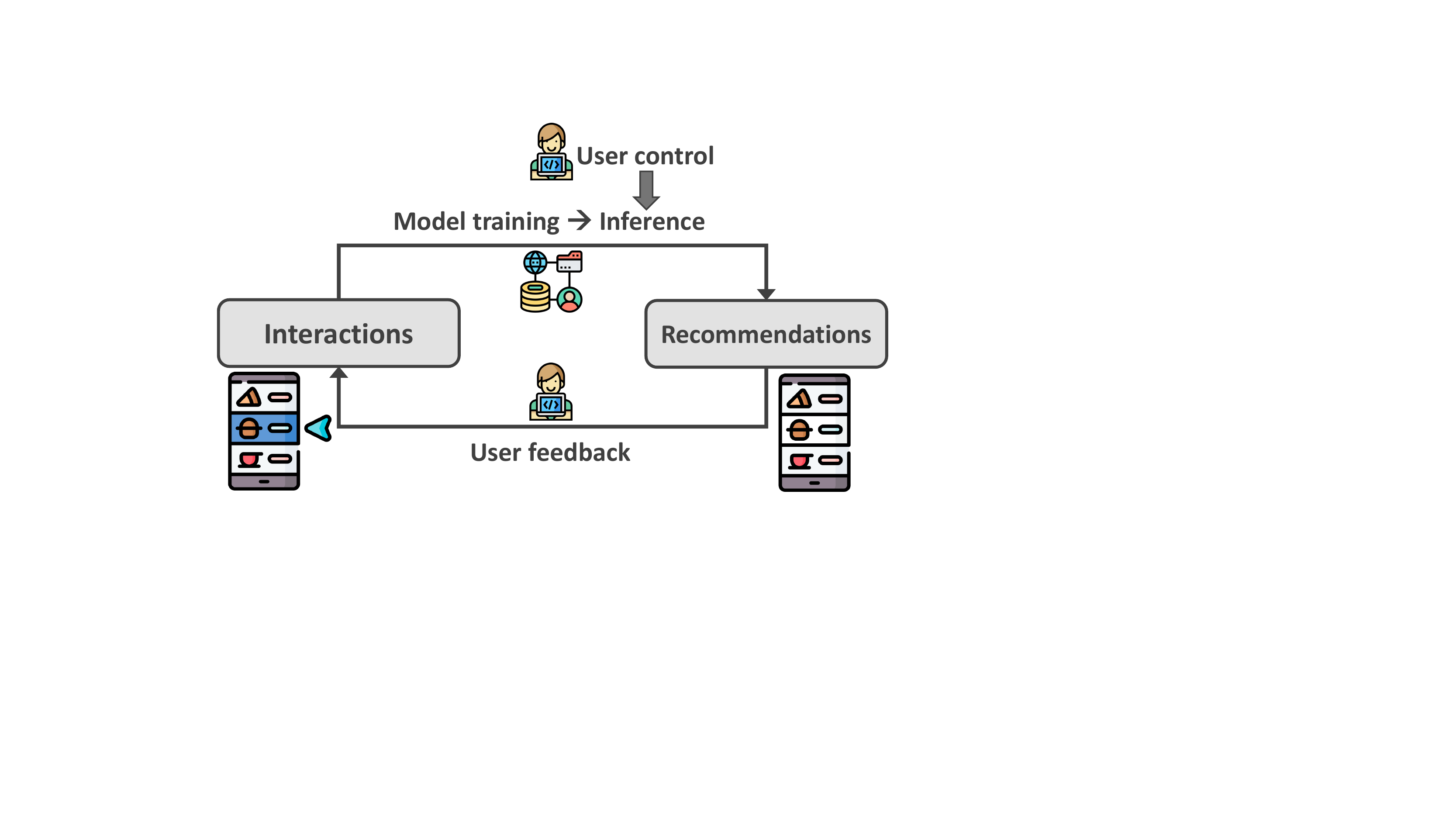}
\caption{Illustration of the proposed UCRS, which introduces another loop between users and recommender systems for the detection of filter bubbles and user controls.}
\label{fig:UCRS}
\end{figure}

\subsection{Formulation of UCRS}

\subsubsection{\textbf{User-controllable Recommender Systems}}

As shown in Figure \ref{fig:UCRS}, UCRS introduces another loop between the users and recommender systems by incorporating two modules: \textit{detection and control modules}. First, the detection module is used to measure the severity of filter bubbles over time, and alert users if they are heavily stuck in filter bubbles. If users are willing to mitigate filter bubbles, they can utilize the control commands and perform real-time adjustments over the recommendations by the control module. 

Formally, given users' historical interactions ${D}$, traditional recommender models aim to predict the recommendations ${R}$ via $P({R}|{D})$. In contrast, UCRS additionally considers user controls $C$ and estimates $P({R}|{D}, do(C))$ with the user interventions $do(C)$~\cite{pearl2009causality}, where the interventions formulate the four kinds of controls from a causal view. 
By the interventions, users are able to quickly adjust the recommendations, significantly decrease the items in historical majority categories, and freely jump out of filter bubbles. 
Specifically, we formulate four kinds of user controls based on user and item features at the fine-grained and coarse-grained levels. 

\subsubsection{\textbf{User-feature Controls}}
We represent $N$ features of user $u$ as $\bm{x_u}=\left[x^1_u, ..., x^n_u, ..., x^{N}_u\right]$, where $x^n_u \in \{0,1\}$ denotes whether user $u$ has the feature $x^n$. For instance, if $[x^1, x^2]$ represents the features of male and female, $\bm{x}_u=[0,1]$ indicates that user $u$ is female. 

\vspace{5pt}
\noindent$\bullet$ \textbf{Fine-grained user-feature controls.} 
To alleviate the filter bubbles \wrt user features (\eg gender and age), we design the fine-grained user-feature controls, which prompt UCRS to recommend more items liked by other user groups. For example, 30-year-old users might have interests in the videos liked by teenagers. 
Formally, to perform $P({R}|{D}, do(C))$ for user $u$, we formulate the control as $do\left(C=c_u\left(+\hat{x}, \alpha\right)\right)$, where $c_u\left(+\hat{x}, \alpha\right)$ is the control command to expose more items liked by other user group $\hat{x}$ and $c_u\left(+\hat{x}, \alpha\right)$ requires that user $u$ does not have feature $\hat{x}$, \ie $\hat{x}_u=0$ for user $u$. Besides, $\alpha \in [0, 1]$ is a coefficient to adjust the strength of user controls on recommendations. 

\vspace{5pt}
\noindent$\bullet$ \textbf{Coarse-grained user-feature controls.}
However, users might simply want to mitigate filter bubbles and do not enjoy the items liked by other user groups. In addition, some users possibly do not know which user group is attractive. As such, we propose the coarse-grained user-feature controls, which help to jump out the filter bubbles of users' own groups. For example, 30-year-old users may not wish the recommendations to be restricted by the feature ``age=30''. 
Formally, the control $do(C)$ in $P({R}|{D}, do(C))$ is formulated as $do\left(C=c_u\left(-\bar{x}, \alpha\right)\right)$, which reduces the items liked by user's own group $\bar{x}$, \ie $\bar{x}_u=1$ for user $u$.

\subsubsection{\textbf{Item-feature Controls}}
Although user-feature controls are able to mitigate the filter bubbles \wrt user features, they ignore the filter bubbles caused by user interactions. As shown in Figure \ref{fig:preliminary}(d), recommender models typically expose more items in the historical majority categories. 
Therefore, to complement user-feature controls, we develop item-feature controls to adjust recommendations \wrt item features. Similar to user features, we represent $M$ features of item $i$ as $\bm{h_i}=\left[h^1_i, ..., h^m_i, ..., h^{M}_i\right]$, where $h^m_i \in \{0,1\}$ denotes whether item $i$ has the feature $h^m$, \eg action movies.

\vspace{5pt}
\noindent$\bullet$ \textbf{Fine-grained item-feature controls.} 
If users have the target item categories (\eg more romance movies), fine-grained item-feature controls can be applied to increase their recommendations. Specifically, the intervention $do(C)$ can be expressed as $do(C=c_i(+\hat{h}, \beta))$, where $\hat{h}$ is the target item category and $\beta \in [0,1]$ is to modify the strength of user controls. 

\vspace{5pt}
\noindent$\bullet$ \textbf{Coarse-grained item-feature controls.}
Correspondingly, we suggest coarse-grained item-feature controls to alleviate the users' burden of specifying target item categories. The goal of coarse-grained item-feature controls is to decrease the recommendations of the largest item category in user historical interactions. In particular, the intervention can be denoted as $do(C=c_i(-\bar{h}, \beta))$, where $-\bar{h}$ means to reduce the largest category $\bar{h}$ in the history.

\subsection{Instantiation of UCRS}
The key to instantiating UCRS lies in the implementation of the detection and control modules. 

\subsubsection{\textbf{Detection of Filter Bubbles}}
We suggest several metrics to measure the severity of filter bubbles from different perspectives, such as diversity and isolation. 
At different time periods, we can calculate the metrics and obtain the severity level of filter bubbles (\eg from 1 to 5) by some heuristic rules designed by the recommender platform. Subsequently, the severity level is presented to users and let users decide whether to control filter bubbles. 

\vspace{1pt}
\noindent$\bullet$ {\textbf{Coverage.}}
Filter bubbles usually decrease the diversity of recommended items, and thus we incorporate a widely adopted metric for diversity: coverage, which calculates the number of item categories in the recommendation list~\cite{zheng2021dgcn}. 

\vspace{1pt}
\noindent$\bullet$ {\textbf{Isolation Index.}}
In addition to diversity-based metrics, we propose Isolation Index~\cite{gentzkow2011ideological} to measure the segregation across different user groups, which is popular to estimate the ideological segregation in sociology~\cite{gentzkow2011ideological}. 
Here, we revise it for the recommendation task. 
Formally, given two user groups $a$ and $b$, we can calculate the Isolation Index of their recommendations by 
\begin{equation}
\label{equ:isolation_index}
\begin{aligned}
s = \sum_{i\in \mathcal{I}}\left(\frac{a_i}{a_n}\cdot \frac{a_i}{a_i+b_i}\right) - \sum_{i\in \mathcal{I}}\left(\frac{b_i}{b_{n}}\cdot \frac{a_i}{a_i+b_i}\right),
\end{aligned}
\end{equation}
where $\mathcal{I}$ is the item set; $a_i$ and $b_i$ are the numbers of users in group $a$ and $b$ who receive the recommendation of item $i$. Besides, $a_n = \sum_{i\in \mathcal{I}}a_i$, which is the total frequency of items exposed to the users in group $a$. Meanwhile, $b_n$ is similar to $a_n$. Finally, $s \in [0,1]$ is equal to the weighted average item exposure of group $a$ minus that of group $b$, where the weights are $\frac{a_i}{a_i+b_i}$~\cite{gentzkow2011ideological}. Intuitively, $s$ captures the extent of recommendation segregation between two groups and higher values denote severer segregation. If there are multiple user groups, we take the average value of $s$ between any pair of groups. 

\vspace{1pt}
\noindent$\bullet$ {\textbf{Majority Category Domination} (MCD).}
Isolation index is more suitable to measure the group segregation \wrt user features (\eg age and gender). As to filter bubbles \wrt item features, we can utilize MCD to obtain the proportion of the historically largest item category in recommendation lists. The increase of MCD across different time periods reflects that the filter bubbles \wrt item category are becoming increasingly serious.

\begin{figure}[t]
\setlength{\abovecaptionskip}{0cm}
\setlength{\belowcaptionskip}{-0.4cm}
\centering
\includegraphics[scale=0.58]{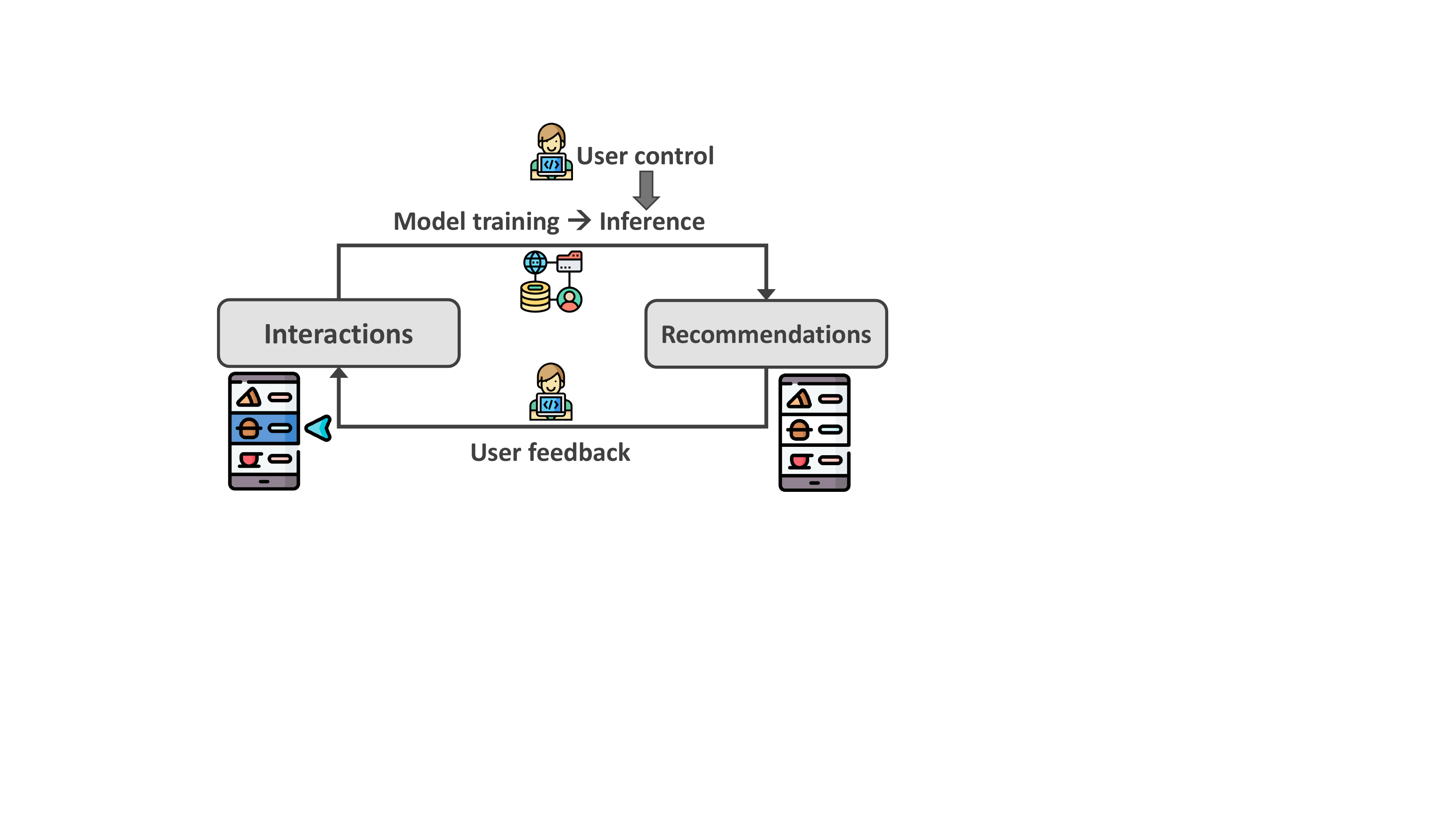}
\caption{Causal graph behind the generation procedure of recommendations.}
\label{fig:causal_graph}
\end{figure}

\subsubsection{\textbf{Response to User-feature Controls}}\label{sec:user_feat_control}

If users aim to mitigate filter bubbles, UCRS needs real-time responses to user controls. For fine-grained user-feature controls $do\left(C=c_u\left(+\hat{x}, \alpha\right)\right)$, users want more items liked by other user group $\hat{x}$. As such, UCRS is actually required to generate recommendations based on the changed user features, which are contrary to the facts. For example, age changes from 30 to 18.
From a causal view, the objective of fine-grained user-feature controls is to answer a counterfactual question~\cite{pearl2009causality}: \textit{what the recommendations \textbf{would} be if the user \textbf{were} in a counterfactual group $\hat{x}$?} 
Similarly, coarse-grained user-feature controls are to answer the question: \textit{what the recommendations \textbf{would} be if the user \textbf{were not} in the factual group $\bar{x}$?} 
To answer the counterfactual questions, the UCI framework needs to inspect the causal relations between user features and recommendations, and then conduct counterfactual inference~\cite{pearl2009causality}. 

\vspace{5pt}
\noindent$\bullet$ \textbf{Causal view of generating recommendations.}
As illustrated in Figure \ref{fig:causal_graph}, we analyze the generation procedure of recommendations by a causal graph~\cite{pearl2009causality}. Specifically, for most models (\eg FM~\cite{rendle2010factorization}), recommender training learns the user representations from interactions, including the representations of ID, age, and gender. Thereafter, the representations of user $u$ and item $i$ are used to predict the probability of user $u$ preferring item $i$, \ie $Y_{u,i} \in [0, 1]$. In detail, $Y_{u,i}$ is fused from the preference scores of the individual ID and multiple group features, where the group preference is shared by the users in the corresponding user group. 

To answer the counterfactual question of fine-grained user-feature controls, an intuitive solution is to change the user features for recommender inference, \eg changing the age from 30 to 18.
As to coarse-grained user-feature controls, we can directly discard the user feature $\bar{x}$ (\eg the age of 30) for inference. 
However, as shown in Figure \ref{fig:causal_graph}, user interactions are actually confounders, which affect the representations of user ID and other group features (\eg age) during recommender training. Therefore, the correlations exist between the representations of user ID and group features. Although the group features are changed or discarded, user ID representations still encode the out-of-date interests of original features, which are inconsistent with user controls and hinder the recommendations of target user groups. 

To remove the confounding effect, the popular choices are confounder balancing~\cite{rubin2005causal}, back-door and front-door adjustments~\cite{pearl2009causality}. Nevertheless, confounder balancing and back-door adjustment require estimating the causal effect of confounders on representations. The estimation is infeasible because 1) user interactions are in a dynamic high-dimension space where new interactions are continually increasing; and 2) the effect of user interactions on representations is decided by the recommender training process, which differs across models and training manners (\eg optimizer and learning rate). Besides, front-door adjustment needs to discover the mediator that blocks all back-door paths, which is not applicable in the causal graph of Figure \ref{fig:causal_graph}. To avoid these challenges, we propose to directly reduce the causal effect of user ID representations on the prediction $Y_{u,i}$ during inference, which can effectively decrease the influence of out-of-date representations without knowing the training process. 

\vspace{5pt}
\noindent$\bullet$ \textbf{Implementation of counterfactual inference.}
In particular, the UCI framework first estimates the effect of user ID representations via counterfactual inference, and then deducts it from the original prediction $Y_{u,i}$~\cite{wang2021click, feng2020graph, Feng2021Empowering}. 
Formally, we image \textit{what the prediction $Y_{\hat{u},i}$ \textbf{would} be if user $u$ \textbf{had not} the ID representations} in a counterfactual world~\cite{wang2021click}, where $\hat{u}$ denotes the representations of user $u$ without ID representations. By comparing $Y_{{u},i}$ with $Y_{\hat{u},i}$, we can measure the effect of user ID representations by $Y_{{u},i} - Y_{\hat{u},i}$. Thereafter, we subtract it from the original prediction $Y_{{u},i}$ with the coefficient $\alpha$:
\begin{equation}
\label{equ:CI}
\small
\begin{aligned}
&Y_{{u},i} - \alpha\cdot \left(Y_{{u},i} - Y_{\hat{u},i}\right) \\
&=f(u,i) - \alpha\cdot \left(f(u,i) - f\left(\hat{u}, i\right)\right) \\
&=(1-\alpha)\cdot f(u,i) + \alpha \cdot f(\hat{u}, i),
\end{aligned}
\end{equation}
where $f(\cdot)$ can be any recommender function of using user and item representations to calculate the prediction $Y$ (\eg FM), and $\alpha\in [0,1]$ adjusts the strength of mitigating the effect of user ID representations. 

\vspace{-6pt}
\begin{center}
\fcolorbox{black}{gray!6}{\parbox{0.98\linewidth}{\noindent$\bullet$ \textbf{Summary of UCI.} The UCI framework consists of two steps to answer the two questions of user-feature controls during inference: 1) changing specific user features to $\hat{x}$ for fine-grained controls and discarding the user feature $\bar{x}$ for coarse-grained controls; and 2) using counterfactual inference to mitigate the effect of out-of-date user ID representations via Equation (\ref{equ:CI}).}}
\end{center}

\subsubsection{\textbf{Response to Item-feature Controls}}

As to item-feature controls, fine-grained controls aim to increase the target item category $\hat{h}$ while coarse-grained ones are to decrease the largest item category $\bar{h}$ of users' history. Indeed, they are asking two interventional questions~\cite{pearl2009causality}: \textit{what the recommendations \textbf{will} be if users \textbf{want} more items in the target category $\hat{h}$ or users \textbf{do not want} the largest category $\bar{h}$?} To answer such questions, the UCI framework utilizes a user-controllable ranking policy as:
\begin{equation}
\label{equ:reranking}
\begin{aligned}
&Y'_{{u},i} = Y_{{u},i} + \beta \cdot r(i), 
\end{aligned}
\end{equation}
where $Y'_{{u},i}$ is the revised score for ranking, and $\beta \in [0, 1]$ is a coefficient to adjust the strength of user controls. Besides, $r(i)$ denotes a regularization term over item $i$. Specifically,
\begin{equation}
\label{equ:reg}
\small
r(i) = \left\{ 
\begin{aligned}
2,  & \quad\text{ if $\hat{h}_i = 1$ for item $i$ with fine-grained controls} \\
0, & \quad\text{ if $\bar{h}_i = 1$ for item $i$ with coarse-grained controls} \\
1,  & \quad\text{ otherwise}, \\
\end{aligned}
\right. 
\end{equation}
where $r(i)$ encourages more recommendations of the items in the target category $\hat{h}$ under fine-grained controls, and decreases the largest category $\bar{h}$ if coarse-grained controls are applied.

\vspace{5pt}
\noindent$\bullet$ \textbf{Target category prediction.}
Due to the extensive item categories, it is a burden for users to specify target categories in fine-grained item-feature controls. Although coarse-grained controls partly alleviate the burden, we can further enhance it by predicting the possible target categories for users. Specifically, if users wish to decrease the largest category of the history, we can predict which item category users will prefer, and then improve the coarse-grained item-feature controls with fine-grained ones. 

\begin{figure}[t]
\setlength{\abovecaptionskip}{0.1cm}
\setlength{\belowcaptionskip}{0cm}
\centering
\includegraphics[scale=0.6]{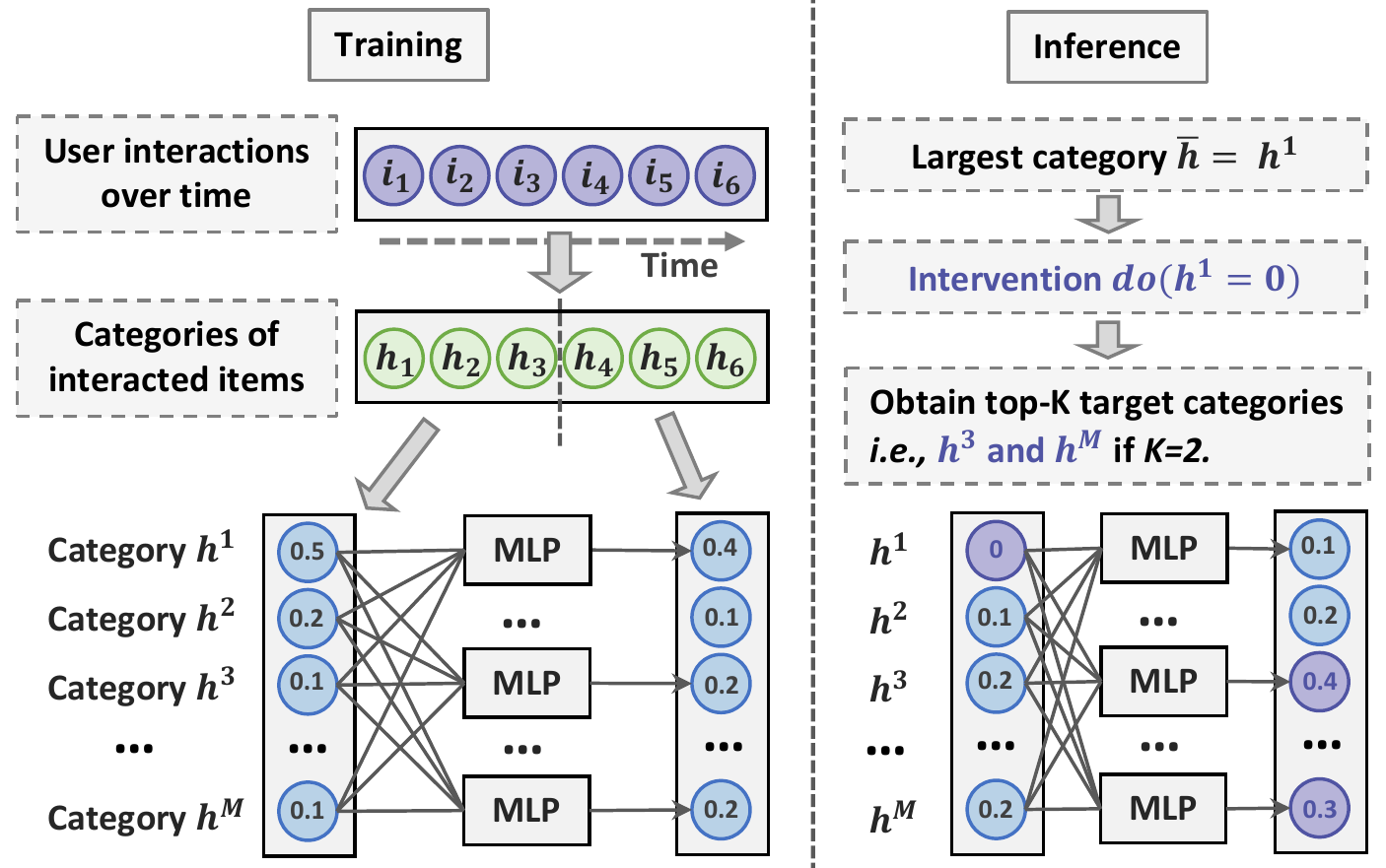}
\caption{Illustration of the training and inference procedures to predict the target item categories.}
\label{fig:target_pred}
\end{figure}

As shown in Figure \ref{fig:target_pred}, we sort users' interacted items by time, and then split the interaction sequence into two parts to obtain the distributions over item categories, respectively. Next, we predict the second category distribution based on the first one via multiple Multi-Layer Perceptrons (MLPs). During training, MLPs utilize the category distributions of all users to capture 1) the temporal interest transition (\eg the increasing preference over some categories), and 2) the relationships between item categories (\eg the users liking action movies probably prefer crime movies). In the inference stage, we leverage the second category distribution to predict top-$K$ target categories. Besides, we conduct intervention $do(\bar{h} = 0)$ to indicate the user controls of reducing category $\bar{h}$. The top-$K$ item categories with the highest values are treated as the target ones in the fine-grained controls. Finally, UCI further enhances coarse-grained item-feature controls by using target category prediction.

\vspace{-6pt}
\begin{center}
\fcolorbox{black}{gray!6}{\parbox{0.98\linewidth}{\noindent$\bullet$ \textbf{Summary of UCI.} Under item-feature controls, user ID representations also encode the historical interests, which conflict with the objective of increasing target categories or decreasing the historical majority category. 
Therefore, 1) UCI first conducts counterfactual inference to mitigate the causal effect of user ID representations on $Y_{u,i}$ as illustrated in Equation (\ref{equ:CI}); 2) for coarse-grained item-feature controls, UCI leverages target category prediction to obtain the top-$K$ target categories; and 3) UCI adopts the ranking policy in Equation (\ref{equ:reranking}) for recommendations.}}
\end{center}

\section{Experiments}
\label{sec:experiment}

We conduct experiments to answer the following questions:
\begin{itemize}[leftmargin=*]
    \item \textbf{RQ1.} How does UCI perform to adjust recommendations for alleviating filter bubbles via four kinds of user controls?
    \item \textbf{RQ2.} How do users can control the recommendations by the coefficients (\ie $\alpha$ and $\beta$)?
    \item \textbf{RQ3.} How does the proposed counterfactual inference affect the recommendations?
\end{itemize}

\begin{table}[t]
\setlength{\abovecaptionskip}{0cm}
\setlength{\belowcaptionskip}{-0.2cm}
\caption{Statistics of the three datasets. ``\#IC'' denotes the number of item categories.}
\label{tab:statistics}
\begin{center}
\resizebox{0.48\textwidth}{!}{
\begin{tabular}{l|l|l|l|l|l}
\hline
Datasets & \#Users & \#Items & \#Interactions & Density & \#IC \\ \hline \hline
DIGIX-Video & 7,643 & 15,526 & 316,045 & 0.0027 & 135 \\ \hline
ML-1M & 6,040 & 3,883 & 575,276 & 0.0245 & 18 \\ \hline
Amazon-Book & 29,115 & 16,845 & 1,712,409 & 0.0035 & 29 \\ \hline
\end{tabular}
}
\end{center}
\end{table}

\subsection{Experimental Settings}
\vspace{5pt}
\noindent\textbf{$\bullet$ Datasets.}
We utilize three real-world datasets for experiments: DIGIX-Video, ML-1M, and Amazon-Book, which are publicly available and vary in terms of domain, user/item features, and sparsity. The statistics of the datasets are presented in Table \ref{tab:statistics}. Specifically,
1) DIGIX-Video\footnote{\url{https://www.kaggle.com/voler2333/2021-digix-video-recommendation.}} is a video recommendation dataset, released by 2021 DIGIX AI Challenge. It covers rich user and item features, including age, gender, and item category. 
2) ML-1M\footnote{\url{https://grouplens.org/datasets/movielens/1m/.}} is a widely-used movie dataset, in which each movie usually has multiple categories.
3) Amazon-Book\footnote{\url{https://nijianmo.github.io/amazon/index.html.}} is a popular dataset for book recommendations, where users only have ID features and each item has a hierarchical taxonomy. In contrast to multi-label item categories in ML-1M, we only keep the largest category, and thus each book is assigned with only one category. 
For each dataset, we use the 10-core setting and treat the interactions with ratings $\geq 4$ as positive samples. In addition, we sort interactions by timestamps, and then split 80\%, 10\%, and 10\% of interactions as the training, validation, and test sets, respectively. For each interaction, we randomly sample an unobserved interaction as the negative sample for training.

\vspace{5pt}
\noindent\textbf{$\bullet$ Evaluation of user-feature controls.}
Since online testing is expensive and infeasible for researchers, we design an offline evaluation setting: 1) assuming some users are willing to mitigate filter bubbles and provide the four kinds of controls; 2) generating the recommendations by different recommender methods according to user controls; and 3) evaluating the recommendations in terms of accuracy and the metrics on mitigating filter bubbles, such as Isolation Index, MCD, and Coverage. 

\textbf{Datasets.}
For user-feature controls, we utilize DIGIX-Video for evaluation because it has rich user features (\ie gender and age) and video categories. In contrast, Amazon-Book only has user ID features; and the users in ML-1M are heavily affected by the dataset bias, where the favorite movie categories of over 77\% of users are ``drama'' and ``comedy'', and 10\% popular movies occupy 52\% interactions. Consequently, the users with different features (\eg age) show similar interaction distributions. As such, Amazon-Book and ML-1M are not well suitable for evaluating the filter bubbles \wrt user features. In this work, we test the fine-grained and coarse-grained user-feature controls over the gender groups and the age groups of DIGIX-Video, respectively. The users take the opposite gender group as the target under fine-grained controls and want to jump out of their own age groups under coarse-grained controls.
This is because the number of age groups is larger, and thus users are more likely to utilize coarse-grained controls without the burden of specifying target age groups.

\vspace{2pt}
\textbf{Baselines.}
All the baselines and our proposed UCI are model-agnostic, which are compared on two representative recommender models: FM and Neural Factorization Machine (NFM)~\cite{he2017nfm}. 
\begin{itemize}[leftmargin=*]
    \item [1)] \textbf{woUF} trains the models without user features (woUF) such as age and gender, which possibly alleviate the segregation across user groups during recommender training.
    \item [2)] \textbf{changeUF} utilizes well-trained recommender models and only changes user features (UF) to the target $\hat{x}$ for inference, \eg changing age from 30 to 18. ChangeUF is used for the fine-grained user-feature controls. 
    \item [3)] \textbf{maskUF} discards the original user features $\bar{x}$ (\eg age=30) for the inference of coarse-grained user-feature controls.
    \item [4)] \textbf{Fairco}~\cite{Morik2020Controlling} is a user-controllable ranking algorithm, which pursues fair exposure opportunities across item groups. \item [5)] \textbf{Diversity}~\cite{ziegler2005improving} incorporates a re-ranking method to diversify recommendations by minimizing the intra-list similarity.
\end{itemize}

\textbf{Metrics.}
To measure the performance, we utilize the all-ranking protocol~\cite{wang2021click}, and the top-10 items are returned as recommendations. We adopt \textbf{Recall} and \textbf{NDCG} to evaluate the accuracy. To quantify the severity of filter bubbles, we leverage \textbf{Isolation Index} (Iso-Index) and \textbf{Coverage} to estimate the group segregation and diversity. In addition, for the fine-grained user-feature controls with target user groups, we develop the metrics \textbf{DIS-EUC} to compare the distance between the recommendations of users and groups. Formally, we denote $\bar{x}$ and $\hat{x}$ as the original and target groups of user $u$, respectively; $\bm{d}_u \in \mathbb{R}^M$ is the distribution over item categories in the recommendations of user $u$; $\bar{g}_u \in \mathbb{R}^M$ denotes the same distribution by averaging over the users in the original group $\bar{x}$ (\eg 30-year-old users); and $\hat{g}_u \in \mathbb{R}^M$ represents the same distribution of the target group $\hat{x}$. Thereafter, we calculate $\text{DIS-EUC} = \text{dis}(d_u, \hat{g}_u) - \text{dis}(d_u, \bar{g}_u)$ for user $u$, where $\text{dis}(\cdot)$ uses Euclidean distance. DIS-EUC measures the distance difference from the user to two groups, where larger distances indicate severer group segregation and filter bubbles.

\begin{table}[t]
\setlength{\abovecaptionskip}{0cm}
\setlength{\belowcaptionskip}{0cm}
\caption{Performance comparison between UCI and the baselines under the coarse-grained user-feature controls.}
\label{tab:user_coarse}
\begin{center}
\setlength{\tabcolsep}{1.8mm}{
\resizebox{0.45\textwidth}{!}{
\begin{tabular}{l|cccc}
\hline
& Recall $\uparrow$ & NDCG $\uparrow$ & Iso-Index $\downarrow$ & Coverage $\uparrow$ \\ \hline
Random & 0.0008 & 0.0005 & 0.0008 & 11.6185 \\ \hline
FM & {\ul 0.0758} & {\ul 0.0584} & 0.1082 & 9.5191 \\ 
FM-woUF & 0.0757 & 0.0582 & 0.1195 & {\ul 9.9211} \\
FM-maskUF & 0.0756 & 0.0577 & 0.1048 & 9.6185 \\
FM-Fairco & 0.0728 & 0.0534 & 0.1050 & \textbf{9.9241} \\
FM-Diversity & 0.0756 & 0.0574 & {\ul 0.1025} & 9.8742 \\ \hline
FM-UCI & \textbf{0.0767} & \textbf{0.0592} & \textbf{0.0777} & 9.8802 \\ \hline \hline
NFM & \textbf{0.0774} & {\ul 0.0585} & 0.1144 & 10.2670 \\
NFM-woUF & 0.0722 & 0.0542 & 0.1191 & 10.1445 \\
NFM-maskUF & 0.0759 & 0.0575 & 0.1378 & 9.9740 \\
NFM-Fairco & 0.0755 & {0.0571} & 0.1130 & \textbf{10.3458} \\
NFM-Diversity & 0.0741 & 0.0562 & {\ul 0.1026} & {\ul 10.3268} \\ \hline
NFM-UCI & {\ul 0.0767} & \textbf{0.0596} & \textbf{0.0760} & 9.9000 \\ \hline
\end{tabular}
}
}
\end{center}
\vspace{-0.5cm}
\end{table}

\begin{table*}[t]
\setlength{\abovecaptionskip}{0cm}
\setlength{\belowcaptionskip}{0cm}
\caption{Performance comparison between UCI and the baseline under the fine-grained user-feature controls. The best results are highlighted in bold and the second best ones are underlined.}
\label{tab:user_fine}
\begin{center}
\setlength{\tabcolsep}{2mm}{
\resizebox{\textwidth}{!}{
\begin{tabular}{l|cc|cc|c|cc|cc|c}
\hline
& \multicolumn{5}{c|}{FM} & \multicolumn{5}{c}{NFM} \\ \hline
& Recall $\uparrow$ & NDCG $\uparrow$ & {Iso-Index} $\downarrow$ & DIS-EUC $\downarrow$ & Coverage $\uparrow$ & Recall $\uparrow$ & NDCG $\uparrow$ & {Iso-Index} $\downarrow$ & DIS-EUC $\downarrow$ & Coverage $\uparrow$ \\ \hline
Random & 0.0008 & 0.0005 & 0.0008 & 0.0001 & 11.6185 & 0.0008 & 0.0005 & 0.0008 & 0.0001 & 11.6185 \\ \hline
FM/NFM & {\ul 0.0861} & {\ul 0.0650} & 0.1161 & 0.0568 & 8.6781 & \textbf{0.0849} & {\ul 0.0630} & 0.1046 & 0.0436 & 9.5126 \\ 
woUF & 0.0858 & 0.0648 & 0.1156 & 0.0561 & 8.7526 & {\ul 0.0847} & 0.0630 & 0.1048 & 0.0431 & 9.5193 \\
changeUF & 0.0858 & 0.0649 & 0.1152 & 0.0566 & 8.6859 & 0.0839 & 0.0626 & 0.1035 & 0.0432 & 9.5461 \\ 
Fairco & 0.0782 & 0.0550 & {0.1082} & {\ul 0.0533} & {\ul 9.1206} & 0.0619 & 0.0420 & 0.1011 & {\ul 0.0357} & {\ul 9.7353} \\
Diversity & 0.0750 & 0.0573 & {\ul 0.0995} & 0.0552 & \textbf{ 9.4312} & 0.0731 & 0.0552 & {\ul 0.0864} & 0.0399 & \textbf{9.8614} \\ \hline
UCI & \textbf{0.0870} & \textbf{0.0661} & \textbf{0.0979} & \textbf{0.0516} & 9.0304 & 0.0844 & \textbf{0.0635} & \textbf{0.0844} & \textbf{0.0354} & 9.6439 \\ \hline
\end{tabular}
}
}
\end{center}
\end{table*}

\begin{table*}[t]
\setlength{\abovecaptionskip}{0cm}
\setlength{\belowcaptionskip}{0cm}
\caption{Results of item-feature controls on ML-1M and Amazon-Book. ``UB'' implies that F-UCI is the ``upper bound'' of C-UCI.}
\label{tab:item_fine}
\begin{center}
\setlength{\tabcolsep}{1mm}{
\resizebox{\textwidth}{!}{
\begin{tabular}{l|cccccc|cccccc}
\hline
& \multicolumn{6}{c|}{ML-1M} & \multicolumn{6}{c}{Amazon-Book} \\ \hline
Method & Recall $\uparrow$ & NDCG $\uparrow$ & W-NDCG $\uparrow$ & MCD $\downarrow$ & TCD $\uparrow$ & Coverage $\uparrow$ & Recall $\uparrow$ & NDCG $\uparrow$ & W-NDCG $\uparrow$ & MCD $\downarrow$ & TCD $\uparrow$ & Coverage $\uparrow$ \\ \hline
Random & 0.0029 & 0.0031 & 0.0029 & 0.2859 & 0.0446 & 8.5024 & 0.0004 & 0.0004 & 0.0004 & 0.2414 & 0.0163 & 4.5892 \\ \hline
FM & 0.0659 & 0.0536 & 0.0485 & 0.5994 & 0.2222 & 8.6600 & 0.0118 & 0.0095 & 0.0085 & 0.5353 & 0.2305 & 3.0175 \\
FM-woIF & 0.0649 & 0.0529 & 0.0481 & 0.5952 & 0.2238 & 8.6755 & 0.0116 & 0.0097 & 0.0084 & 0.5310 & 0.2268 & 3.1792 \\
FM-Fairco & 0.0605 & 0.0506 & 0.0459 & 0.3812 & 0.2417 & {\ul 9.3799} & 0.0117 & 0.0095 & 0.0085 & 0.1559 & 0.2942 & \textbf{3.2700} \\
FM-Diversity & 0.0531 & 0.0473 & 0.0428 & 0.5597 & 0.2351 & \textbf{9.5407} & 0.0092 & 0.0080 & 0.0072 & 0.5199 & 0.2383 & {\ul 3.1899} \\ \hline
FM-Reranking & 0.0761 & 0.0637 & 0.0603 & \textbf{0.1000} & 0.3099 & 8.9409 & {\ul 0.0178} & {\ul 0.0142} & 0.0142 & {\ul 0.0026} & 0.5348 & 3.0196 \\
FM-C-UCI & {\ul 0.0770} & {\ul 0.0665} & {\ul 0.0630} & {\ul 0.2466} & {\ul 0.3334} & 9.1206 & 0.0173 & 0.0141 & {\ul 0.0146} & 0.0213 & {\ul 0.6310} & 2.0768 \\
FM-F-UCI (UB) & \textbf{0.2095} & \textbf{0.1704} & \textbf{0.1792} & 0.3544 & \textbf{1.0000} & 8.0712 & \textbf{0.0334} & \textbf{0.0283} & \textbf{0.0337} & \textbf{0.0023} & \textbf{1.0000} & 1.0002 \\ \hline \hline
NFM & 0.0651 & 0.0556 & 0.0501 & 0.5748 & 0.2321 & 8.8854 & 0.0121 & 0.0102 & 0.0088 & 0.5488 & 0.2294 & 2.9818 \\
NFM-woIF & 0.0654 & 0.0551 & 0.0498 & 0.5732 & 0.2290 & 9.0110 & 0.0112 & 0.0092 & 0.0082 & 0.5330 & 0.2332 & 3.0900 \\
NFM-Fairco & 0.0626 & 0.0516 & 0.0470 & 0.4750 & 0.2441 & {\ul 9.3715} & 0.0114 & 0.0091 & 0.0085 & 0.1856 & 0.2891 & \textbf{3.8497} \\
NFM-Diversity & 0.0522 & 0.0481 & 0.0438 & 0.5391 & 0.2391 & \textbf{9.7018} & 0.0109 & 0.0092 & 0.0081 & 0.5146 & 0.2427 & {\ul 3.1825} \\ \hline
NFM-Reranking & 0.0752 & 0.0672 & 0.0631 & \textbf{0.0296} & {\ul 0.3163} & 9.0468 & 0.0180 & 0.0144 & 0.0144 & {\ul 0.0025} & 0.5421 & 3.0184 \\
NFM-C-UCI & {\ul 0.0778} & {\ul 0.0687} & {\ul 0.0647} & {\ul 0.2753} & 0.3119 & 9.0744 & {\ul 0.0181} & {\ul 0.0148} & {\ul 0.0154} & 0.0049 & {\ul 0.6779} & 1.4190 \\
NFM-F-UCI (UB) & \textbf{0.2125} & \textbf{0.1729} & \textbf{0.1820} & 0.3319 & \textbf{1.0000} & 8.2299 & \textbf{0.0338} & \textbf{0.0276} & \textbf{0.0327} & \textbf{0.0023} & \textbf{1.0000} & 1.0002 \\ \hline
\end{tabular}
}
}
\end{center}
\end{table*}

\vspace{3pt}
\noindent\textbf{$\bullet$ Evaluation of item-feature controls.}
We conduct experiments on the users who have preference shifts from the training to test sets. Specifically, for each user, we obtain the largest item categories in the training and test sets, and then we select the users with different largest categories. This simulates the situation that users aim to mitigate historical filter bubbles and want more items in other categories. The numbers of selected users in DIGIX-Video, ML-1M, and Amazon-Book are 4320, 3806, and 5155, respectively. 

\vspace{1pt}
\textbf{Baselines.}
We generate recommendations for the selected users by using the following methods:
1) \textbf{woIF} trains FM and NFM without item features (woIF); 
2) \textbf{Fairco}; 
3) \textbf{Diversity}; 
4) \textbf{Reranking} is one variant of UCI, which only uses the ranking policy in Equation (\ref{equ:reranking}) and discards counterfactual inference and target category prediction; 
5) \textbf{C-UCI} denotes the UCI strategy with target category prediction under coarse-grained controls; and 
6) \textbf{F-UCI} represents UCI under fine-grained controls, which knows the target category of each user, \ie the largest category in the test set. 

\vspace{1pt}
\textbf{Metrics.}
For performance comparison, we use Recall, NDCG, and Coverage. Besides, we introduce a new metric \textbf{Weighted NDCG} (W-NDCG), which assigns the NDCG relevance scores of the positive items in the target categories, the positive ones in non-target categories, and the negative ones as 2, 1, and 0, respectively. W-NDCG distinguishes the positive items in the target and non-target categories, and prefers the positive ones in the target categories. 
Furthermore, we employ MCD and \textbf{Target Category Domination} (TCD) to calculate the proportions of the historical majority category and users' target category in the recommendations, respectively.

\vspace{2pt}
\noindent\textbf{$\bullet$  Hyper-parameter settings.}
We train FM and NFM by following the settings in~\cite{he2017nfm}: the sizes of user/item representations are 64; and Adagrad with the batch size of 1,024 is used for parameter optimization. The learning rate is searched in $\{0.001, 0.01, 0.05\}$. The hidden size of the MLP in NFM and target category prediction is tuned in $\{4,8,16,32\}$ and the normalization coefficient is searched from $\{0,0.1,0.2\}$. $K$ in target category prediction is chosen from $\{1,2,...,5\}$.
Besides, $\alpha$ and $\beta$ in the controls of $c_u(\cdot)$ and $c_i(\cdot)$ are adjusted in $\{0,0.1,...,0.5\}$ and $\{0,0.01,...,0.1\}$, respectively. We select the best model by Recall on the validation set.

\subsection{Performance Comparison}

\subsubsection{\textbf{Performance under User-feature Controls (RQ1)}}

We present the results under the coarse-grained and fine-grained user-feature controls in Table \ref{tab:user_coarse} and Table \ref{tab:user_fine}, respectively. From the two tables, we have the following observations:

\begin{itemize}[leftmargin=*]
    \item The intuitive baselines (\ie woUF, changeUF, and maskUF) slightly decrease the recommendation accuracy and alleviate the group segregation in terms of Isolation Index and DIS-EUC. Meanwhile, diversity rises marginally in most cases. However, the overall performance is quite similar to FM or NFM. This is consistent with the analysis in Section \ref{sec:user_feat_control}: although the user features are discarded for training or changed/masked for inference, the user ID representations still encode the historical interactions, which are affected by the users' original features and lead to similar recommendations with FM and NFM. 
    
    \item Fairness and diversity methods can effectively mitigate the issue of filter bubbles and diversify the recommendation lists. Nevertheless, they bring sharp performance drop. For example, the accuracy of Fairco on FM declines by 15.38\% \wrt NDCG in Table \ref{tab:user_fine}. It makes sense because pursuing the objectives of fairness and diversity will inevitably recommend many irrelevant items, occupying the opportunities of positive items. 
    
    \item UCI significantly alleviates the group segregation in filter bubbles while achieving superior accuracy. Besides, the diversity also increases as compared to FM and NFM, which relieves the accuracy-diversity dilemma. 
    We attribute the improvements to the effectiveness of counterfactual inference in reducing the effect of out-of-date user ID representations. The mitigation of such effect pushes the recommender model to expose fewer items similar to historical interactions, and makes the recommendations more diverse. Meanwhile, the superior accuracy is because $\alpha$ in the user control $c_u(\cdot, \alpha)$ adjusts the influence of the representations of user ID and other group features (\eg age and gender), leading to a better balance between individual preference and group preference as illustrated in Figure \ref{fig:causal_graph}.
\end{itemize}

\begin{figure*}[ht]
\setlength{\abovecaptionskip}{0cm}
\setlength{\belowcaptionskip}{0cm}
  \centering 
  \hspace{-0.7in}
  \subfigure{
    \includegraphics[width=1.48in]{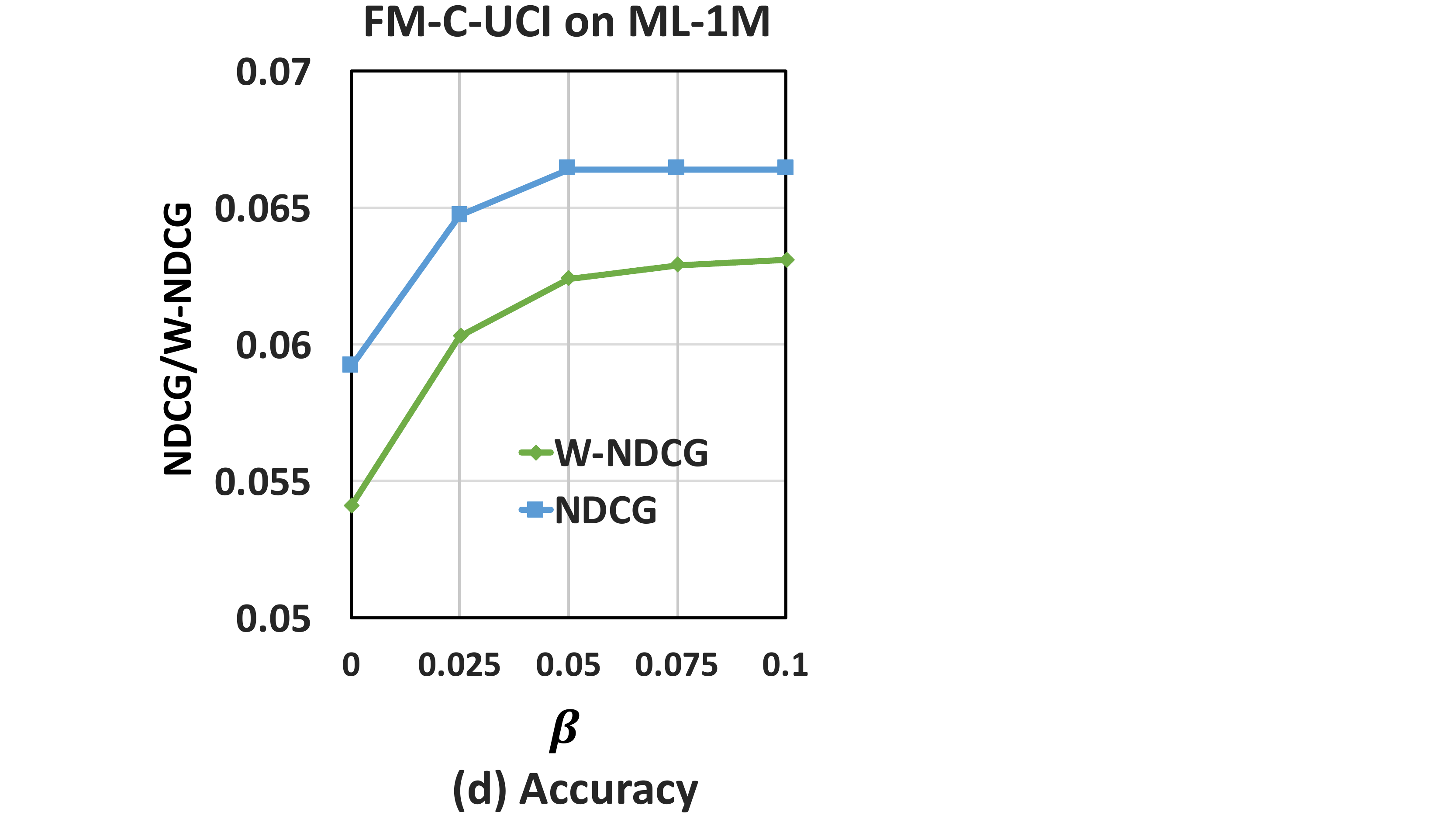}} 
  \hspace{-0.105in}
  \subfigure{
    \includegraphics[width=1.43in]{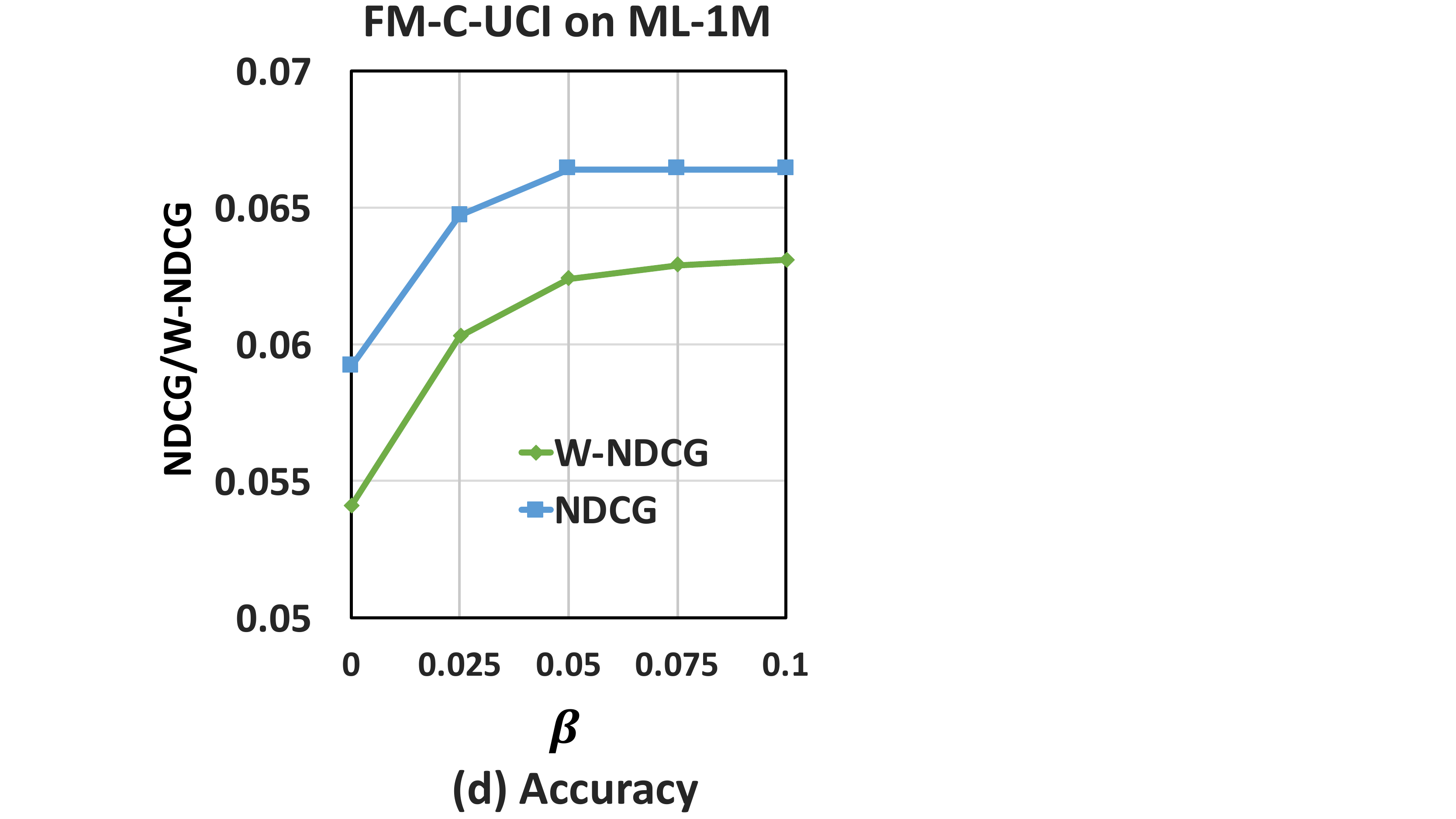}} 
  \hspace{-0.105in}
  \subfigure{
    \includegraphics[width=1.4in]{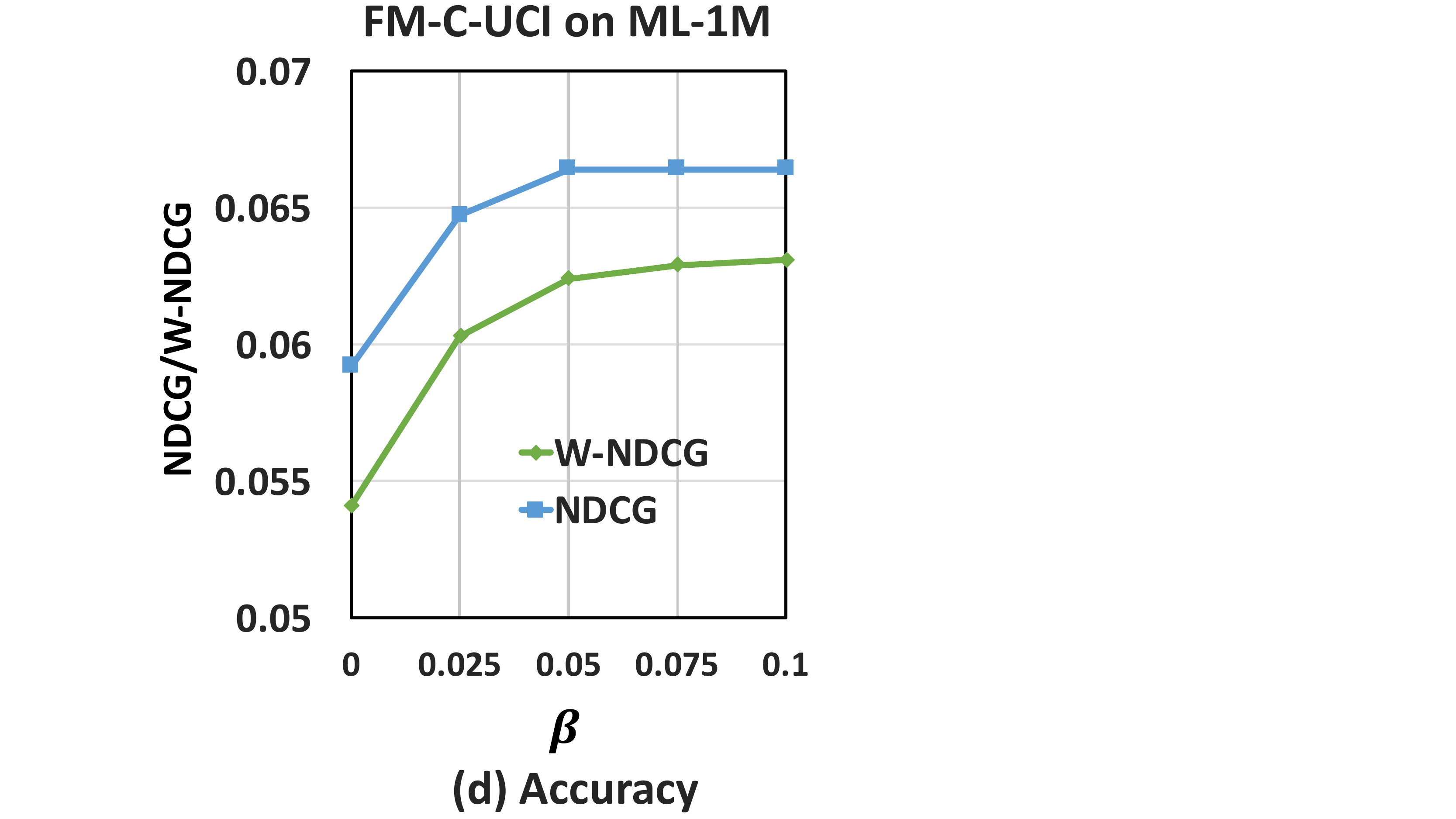}} 
  \hspace{-0.105in}
  \subfigure{
    \includegraphics[width=1.45in]{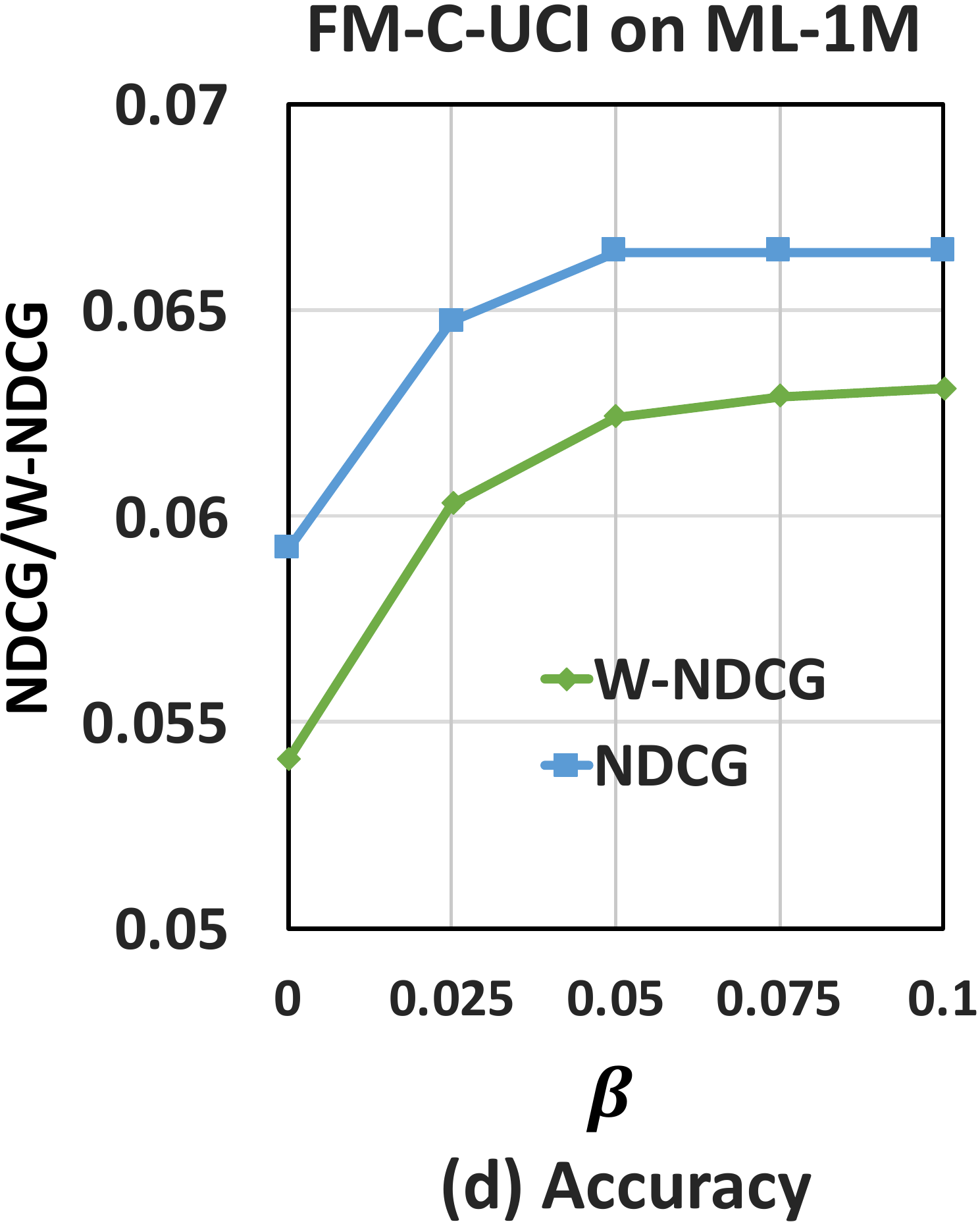}} 
  \hspace{-0.105in}
  \subfigure{
    \includegraphics[width=1.35in]{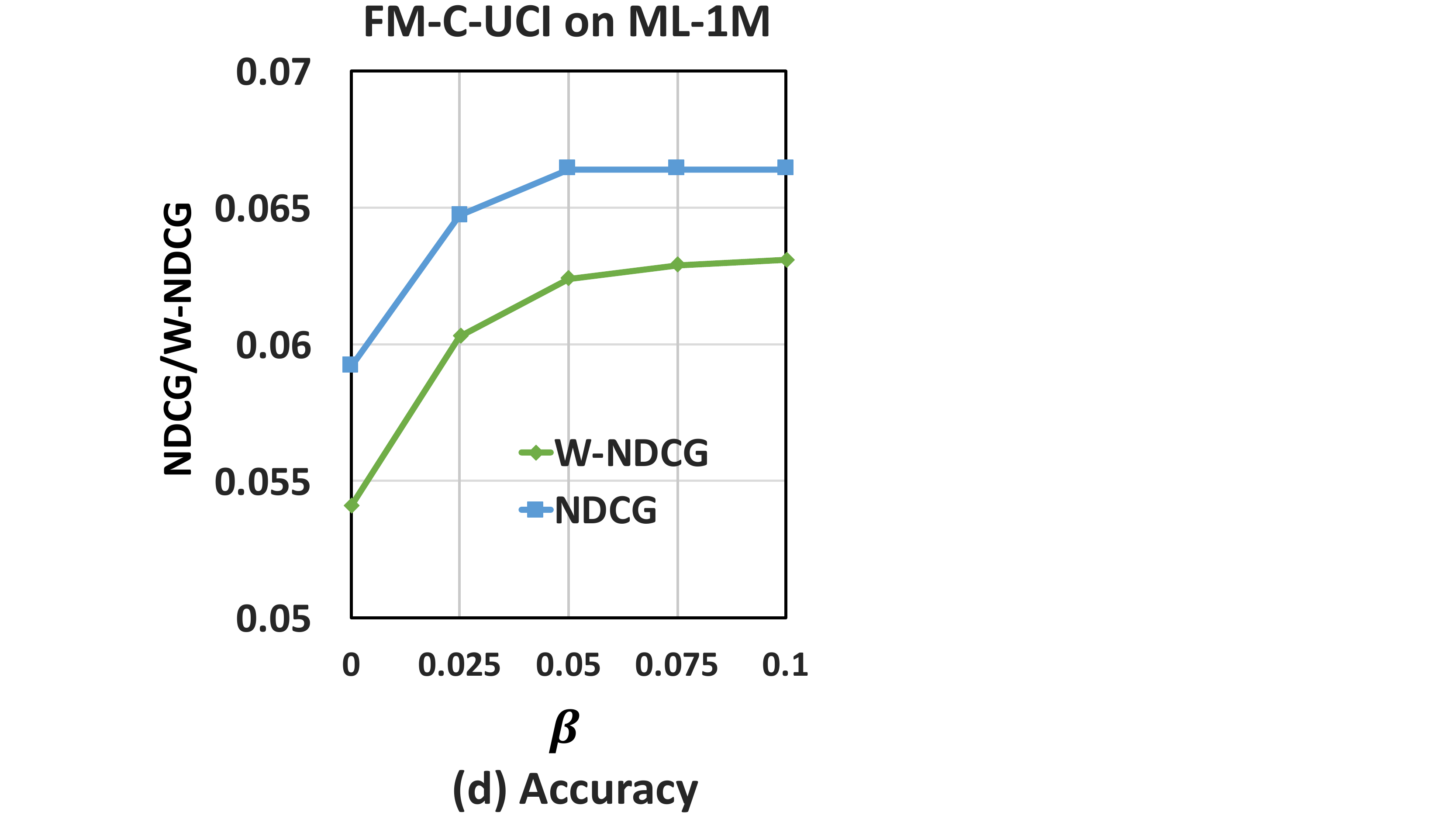}} 
  \hspace{-0.7in} 
  \caption{Effects of the control coefficients $\alpha$ and $\beta$ \wrt accuracy, isolation, diversity, and category domination.} 
  \label{fig:alpha_beta}
\end{figure*}

\subsubsection{\textbf{Performance under Item-feature Controls (RQ1)}}
Table \ref{tab:item_fine} provides the results of item-feature controls on ML-1M and Amazon-Book. The observations on DIGIX-Video are similar to those on ML-1M. The consideration of presenting ML-1M and Amazon-Book is to compare the effects of single-label and multi-label item categories. From Table \ref{tab:item_fine}, we have the following findings:
\begin{itemize}[leftmargin=*]
    \item As compared with FM and NFM, woIF marginally decreases the historical majority categories \wrt MCD, where the marginal effect shows that woIF still recommends many items in the historical majority categories without knowing item features. Moreover, woIF slightly degrades the accuracy and improves the diversity. Such observations and the underlying reasons are analogous to woUF under user-feature controls. 
    
    \item The performance of Fairco and Diversity is similar to that under user-feature controls: at the expense of sacrificing accuracy, they substantially alleviate the filter bubbles by recommending fewer historical majority categories and improving diversity. 
    
    \item Reranking and C-UCI have significant performance improvements over the baselines \wrt the accuracy and mitigation of majority categories. This is due to using coarse-grained item-feature controls, which indicate the largest categories in the history to decrease. Besides, C-UCI performs better than Reranking, especially in terms of W-NDCG, validating the superiority of counterfactual inference and target category prediction. Counterfactual inference reduces the influence of out-of-date ID representations for these users with preference shifts; and meanwhile UCI recommends more target categories due to the target category prediction. Furthermore, as compared to ML-1M, the improvements of UCI over Reranking on Amazon-Book are only significant \wrt W-NDCG. It is reasonable because Amazon-Book only has user ID features and counterfactual inference is impractical: only ID representations are to represent users and reducing their effect usually does not change the ranking lists. Therefore, on Amazon-Book, purely target category prediction is helpful to enhance W-NDCG and TCD. 
    
    \item F-UCI achieves the best accuracy by following users' fine-grained item-feature controls, showing that directly incorporating user controls into recommender inference is greatly effective to understand users' interests as compared to passively learning from user interactions. As the upper bound of C-UCI, it also implies the promising potential of target category prediction, where a more accurate prediction of target categories can lead to dramatic accuracy improvements. 

    \item The performance of diversity varies from ML-1M to Amazon-Book, especially over Reranking, C-UCI, and F-UCI. Both C-UCI and F-UCI decrease the diversity while the relative drop on Amazon-Book is larger. The reasons are that 1) C-UCI and F-UCI emphasize the recommendations of target item categories; and 2) the items in Amazon-Book have single-label categories. Purely recommending the target categories will easily degrade the diversity. This indicates that users should adjust the control coefficients $\alpha$ and $\beta$ to balance the superior accuracy and diversity at their own will. 
\end{itemize}

\subsubsection{\textbf{Effect of Coefficients in User Controls (RQ2)}}

To study whether users can flexibly adjust the recommendations, we explore the effect of the control coefficients $\alpha$ and $\beta$. 
The results of FM-UCI and NFM-UCI \wrt varying $\alpha$ on DIGIX-Video are reported in Figure \ref{fig:alpha_beta}(a), (b), and (c). The performance of FM-C-UCI \wrt $\beta$ on ML-1M is summarized in Figure \ref{fig:alpha_beta}(d) and (e).
More results on other datasets have similar trends, which are omitted to save space. From the figures, we can find: 
1) $\alpha$ controls the influence of user-feature controls, where a larger $\alpha$ significantly alleviates the filter bubbles and enhances the diversity as shown in Figure \ref{fig:alpha_beta}(b) and (c); 2) the accuracy \wrt increasing $\alpha$ first rises, and then gradually decreases. Besides, we can observe that the accuracy drop is more steady as compared to the isolation and diversity. Such findings verify that UCI can mitigate filter bubbles and improve diversity without sacrificing accuracy or with less accuracy decline; and 3) from the results in Figure \ref{fig:alpha_beta}(d) and (e) under item-feature controls, $\beta$ is able to mitigate the domination of historical majority categories while improving the accuracy.

\begin{table}[t]
\setlength{\abovecaptionskip}{0cm}
\setlength{\belowcaptionskip}{0cm}
\caption{Performance comparison with (w/) and without (w/o) counterfactual inference (CI).}
\label{tab:ablation_CI}
\begin{center}
\resizebox{0.48\textwidth}{!}{
\begin{tabular}{l|l|ccccc}
\hline
Method & Variants & Recall & NDCG & W-NDCG & MCD & Coverage \\ \hline
\multirow{2}{*}{FM-F-UCI} & w/o CI & 0.2094 & 0.1689 & 0.1777 & 0.3587 & 7.9519 \\
 & w/ CI & \textbf{0.2095} & \textbf{0.1704} & \textbf{0.1792} & \textbf{0.3544} & \textbf{8.0712} \\ \hline
\multirow{2}{*}{NFM-F-UCI} & w/o CI & 0.2094 & 0.1687 & 0.1775 & 0.3525 & 8.0155 \\
 & w/ CI & \textbf{0.2125} & \textbf{0.1729} & \textbf{0.1820} & \textbf{0.3319} & \textbf{8.2299} \\ \hline
\end{tabular}
}
\end{center}
\vspace{-0.2cm}
\end{table}

\subsubsection{\textbf{Ablation Study of Counterfactual Inference (RQ3)}}
We conduct ablation study to further analyze the effect of counterfactual inference. In Figure \ref{fig:alpha_beta}(a), (b), and (c), $\alpha=0$ denotes the ablation of counterfactual inference under user-feature controls. Besides, we remove it from FM-F-UCI and NFM-F-UCI under item-feature controls, and summarize the results on ML-1M in Table \ref{tab:ablation_CI}. 
From Figure \ref{fig:alpha_beta} and Table \ref{tab:ablation_CI}, we observe that using counterfactual inference alleviates the filter bubbles, improves the diversity, and even enhances the accuracy when $\alpha$ is small. The higher accuracy is mainly due to the inconsistency between out-of-date user ID representations and the latest user interests.

\section{Conclusion and Future Work}
\label{sec:conclusion}

In this work, we proposed a novel recommender prototype UCRS to flexibly alleviate filter bubbles, which provides users more choices to adjust recommendations. 
Functionally, the prototype can detect the severity of filter bubbles and allow users to adjust filter bubbles via user controls. 
In particular, we developed four kinds of user controls: the user-feature and item-feature controls at the fine-grained and coarse-grained levels. 
To implement the user controls, we designed a UCI framework for recommender inference, which leverages counterfactual inference to mitigate the effect of out-of-date user ID representations on recommendations. Furthermore, UCI revises the user features for user-feature controls and adopts a ranking policy with target category prediction for item-feature controls.
We proposed several metrics to measure filter bubbles and conducted experiments on three datasets, validating the effectiveness of UCI in alleviating filter bubbles and maintaining the accuracy. 

The new UCRS prototype and the novel UCI framework can be widely deployed in the practical recommender systems. The recommender platform can design various interactive interfaces (\eg conversational systems and control panels) to acquire the control commands, and then adopt UCRS to effectively adjust recommendations. This additional interaction paradigm between users and recommender systems will 1) ensure the user rights of controlling recommender strategies, 2) increase the user engagement in the recommendation ecosystem, and 3) significantly enhance the user satisfaction over the recommended items.

Nevertheless, this work takes the initial step to perform user-controllable recommendation against filter bubbles, leaving many potential directions to future work. In particular, 
1) it is non-trivial to instantiate the proposed UCRS framework in the online testing platforms, which is costly and impractical for researchers but shows a better justification \wrt the efficiency and effectiveness of UCI;
and 2) more user controls under the framework of UCRS can be designed by collecting users' opinions, which will help users to quickly adjust recommendations by more diverse interfaces.


{
\tiny
\bibliographystyle{ACM-Reference-Format}
\balance
\bibliography{bibfile}
}

\end{document}